\pdfoutput=1
\documentclass[11pt, reqno,preprint]{article}
\usepackage{jheppub}
\usepackage{epsfig}
\usepackage{amssymb}
\usepackage{amsmath}
\usepackage{mathrsfs}
\usepackage{hyperref}
\usepackage{multirow}
\usepackage{bbold}
\usepackage{empheq}

\def\be{\begin{equation}}
\def\ee{\end{equation}}
\def\ba{\begin{eqnarray}}
\def\ea{\end{eqnarray}}
\def\nl{\nonumber\\}

\def\b{\beta}

\def\adj{{\rm adj}}

\def\Li{\textrm{Li}}
\def\l{\langle}
\def\r{\rangle}
\def\SS{\mathcal{S}}

\def\eps{\epsilon}
\def\z#1#2{\alpha_{{#1}{#2}}}
\def\zsq#1#2{\alpha_{{#1}{#2}}^2}

\def\GG{K}
\def\MSbar{\overline{{\rm MS}}}

\def\CA{C_A}

\def\nF{n_F}
\def\nS{n_S}
\def\TF{T_F}
\def\TS{T_S}
\def\TR{T_R}
\def\betazero{b_0}

\def\lorentz{\ell}
\def\gammaK{\gamma_{K}}

\def\ren{{\rm ren}}

\def\Tr{\mbox{Tr}}

\newcommand{\lsim}{\mathrel{\hbox{\rlap{\lower.55ex \hbox{$\sim$}} \kern-.3em \raise.4ex \hbox{$<$}}}}
\newcommand{\gsim}{\mathrel{\hbox{\rlap{\lower.55ex \hbox{$\sim$}} \kern-.3em \raise.4ex \hbox{$>$}}}}

\def\alphas{\alpha_\textrm{s}}

\def\eps{\epsilon}

\title{Resummation of non-global logarithms and\\ the BFKL equation}

\author{Simon Caron-Huot}
\affiliation[a]{Niels Bohr International Academy and Discovery Center, Blegdamsvej 17, Copenhagen 2100, Denmark}
\emailAdd{schuot@nbi.dk} \abstract{
We consider a `color density matrix' in gauge theory.
We argue that it systematically resums large logarithms originating from wide-angle soft radiation, sometimes referred to as non-global logarithms, to all logarithmic orders.  We calculate its anomalous dimension at leading- and next-to-leading order.
Combined with a conformal transformation known to relate this problem to shockwave scattering in the Regge limit,
this is used to rederive the next-to-leading order Balitsky-Fadin-Kuraev-Lipatov equation
(including its nonlinear generalization, the so-called Balitsky-JIMWLK equation),
finding perfect agreement with the literature.  Exponentiation of divergences to all logarithmic orders is demonstrated.
The possibility of obtaining the evolution equation (and BFKL) to three-loop is discussed.
}

\notoc

\begin{document}

\maketitle

\section{Introduction}

Collimated sprays of particles, or jets, figure prominently in high-energy collider physics.
This has led to a growing interest in the characterization of jet shapes and event shapes, with the goal
to extract as much information as possible about underlying hard scattering events.
The pencil-like nature of jets implies that one often encounters disparate angular and energy scales.
These lead to large logarithms in theoretical calculations,
whose resummation is necessary to obtain controlled, precise predictions.
Theoretically, in analytic studies these large logarithms are often the only terms which one may hope to predict
in an amplitude or cross section at higher orders in perturbation theory, and thus could potentially help reveal new structures.
Both of these reasons make them especially important.

Thanks to developments spanning many years, resummation for most observables of interest
is now possible.  In the case of so-called global observables, which involve complete (`global') integrals
over final state phase spaces, the critical ingredient is the exponentiation of
infrared and collinear divergences \cite{Sen:1982bt,Korchemsky:1991zp,Bauer:2000yr,Sterman:2002qn,Aybat:2006mz,Feige:2014wja}.
This predicts in a quantitative way the logarithms left after the cancelation of infrared and collinear divergences,
cancelations which are guaranteed on general grounds by the Kinoshita-Lee-Neuenberg (KLN) theorem \cite{Kinoshita:1962ur,Lee:1964is}.
There exists however non-global observables, for which phase space cuts lead to
soft radiation not being integrated over all angles (`not globally'),
for which large logarithms are considerably more difficult to resum \cite{Dasgupta:2001sh,Banfi:2002hw}.

The aim of this paper is to set up a comprehensive theory of non-global logarithms,
to all logarithmic orders and finite $N_c$, in the cases where collinear singularities are absent.
This theory will turn out to be closely related to that of Balitsky-Fadin-Kuraev-Lipatov (BFKL), which controls large logarithms
in a different limit, the Regge limit (high-energy scattering at fixed momentum transfer) \cite{Kuraev:1977fs,Balitsky:1978ic}.

To set the stage we consider a generic weighted cross-section of the form
\be
 \sigma = \sum_n \int d\Pi_n \big|A_{Q\to n}(p_1,\ldots,p_n)\big|^2 u(\{p_i\}) \label{finite_observable}
\ee
where $d\Pi_n$ is the phases space measure for $n$ partons and
the measurement function $u(\{p_i\})$ specifies the details of the measurement, including various vetoes etc. 
For suitable infrared- and collinear-safe measurements, the cross-section will be finite
order by order in perturbation theory. 
As a preliminary simplification (to avoid initial state radiation), in this paper we will assume that the initial state is a color-singlet state of mass $Q$,
and assume massless final states.

\begin{figure}
\centering
\be
\def\svgwidth{12cm}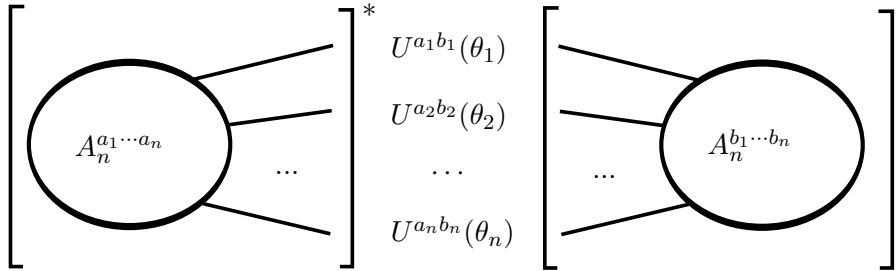
\nonumber\ee
\caption{Color density matrix.
For each colored final state, an independent color rotation is
applied between the amplitude and its complex conjugate.
}\label{fig:densitymatrix}
\end{figure}

A time-tested strategy to resum large logarithms is to introduce intermediate matrix elements
which depend on a factorization scale and use the renormalization group to control the dependence on that scale.
The template is Wilson's operator product expansion, which expresses correlators at short distances in terms of
short-distance OPE coefficients, anomalous dimensions, and long-distance matrix elements. The
factorization scale $\mu$, whose dependence is controlled by the renormalization group,
cancels between the OPE coefficients and matrix elements, thus providing a handle on large logarithms.
Our main proposal is that the pertinent operator for resumming non-global logarithms is the \emph{color density matrix}:
\be
 \sigma[U] \equiv \sum_n \int d\Pi_n
 \Big[A_{Q\to n}^{a_1\cdots a_n}(\{p_i\})\Big]^*
U^{a_1b_1}(\theta_1)\cdots
U^{a_nb_n}(\theta_n) 
 \Big[A_{Q\to n}^{b_1\cdots b_n}(\{p_i\})\Big] u(\{p_i\})\,. \label{density_matrix}
\ee
We call it a density matrix because it is linear in both the amplitude and its complex conjugate.
Note that a full density matrix would allow different momenta in each factor, but here only the color indices are different.
This defines a functional of a continuous field of unitary matrices $U^{ab}(\theta)$,
which depend on a two-dimensional angle and live in the adjoint representation of the gauge group.
Pictorially, $U$, shown in fig.~\ref{fig:densitymatrix}, is a (local) color rotation between the matrix element and its conjugate.  A closely related construction has been used to describe parton showers at finite $N_c$ \cite{Nagy:2012bt}.

The physical motivation for eq.~(\ref{density_matrix}) is that the information carried by $\sigma[U]$
should be necessary and sufficient to fully characterize the distribution soft wide-angle gluons.
Since soft emissions can be triggered by any other colored parton with a higher energy, keeping track of the color flow in every direction, like $\sigma[U]$ does, seems clearly \emph{necessary}.
The information in $\sigma[U]$ is also intuitively \emph{sufficient}: due to coherence effects,
soft gluons are affected by the color charge carried by harder partons but not by other details.

Contrary to the original weighted cross-section, the density matrix $\sigma[U]$ is infrared divergent.
We propose, and will demonstrate, that these infrared divergences exponentiate in terms of a well-defined anomalous dimension operator.
This supports our claim that the information in $\sigma[U]$ is sufficient.
After cancelling these divergences (see eq.~(\ref{factorization})), the renormalized density matrix
then depends on a factorization $\mu$ scale through
\be
  \left[\mu\frac{\partial}{\partial \mu} + \beta \frac{\partial}{\partial\alphas}\right]
   \sigma^{\ren}[U;\mu] = \GG(U,\delta/\delta U,\alphas(\mu),\eps) \,\sigma^{\ren}[U;\mu]\,. \label{RG_for_sigmahard}
\ee
This renormalization group equation then provides the desired handle on large infrared logarithms.
The anomalous dimension operator, or ``Hamiltonian'', $\GG$, assumes the form of a functional differential operator.
Its one-loop expression, given in eq.~(\ref{Gamman_0}) below, reproduces earlier formulas derived in the literature
to deal with non-global logarithms \cite{Banfi:2002hw,Weigert:2003mm}.

\subsection{Structure of the resummation}

\begin{figure}
\centering
\includegraphics[width=6cm]{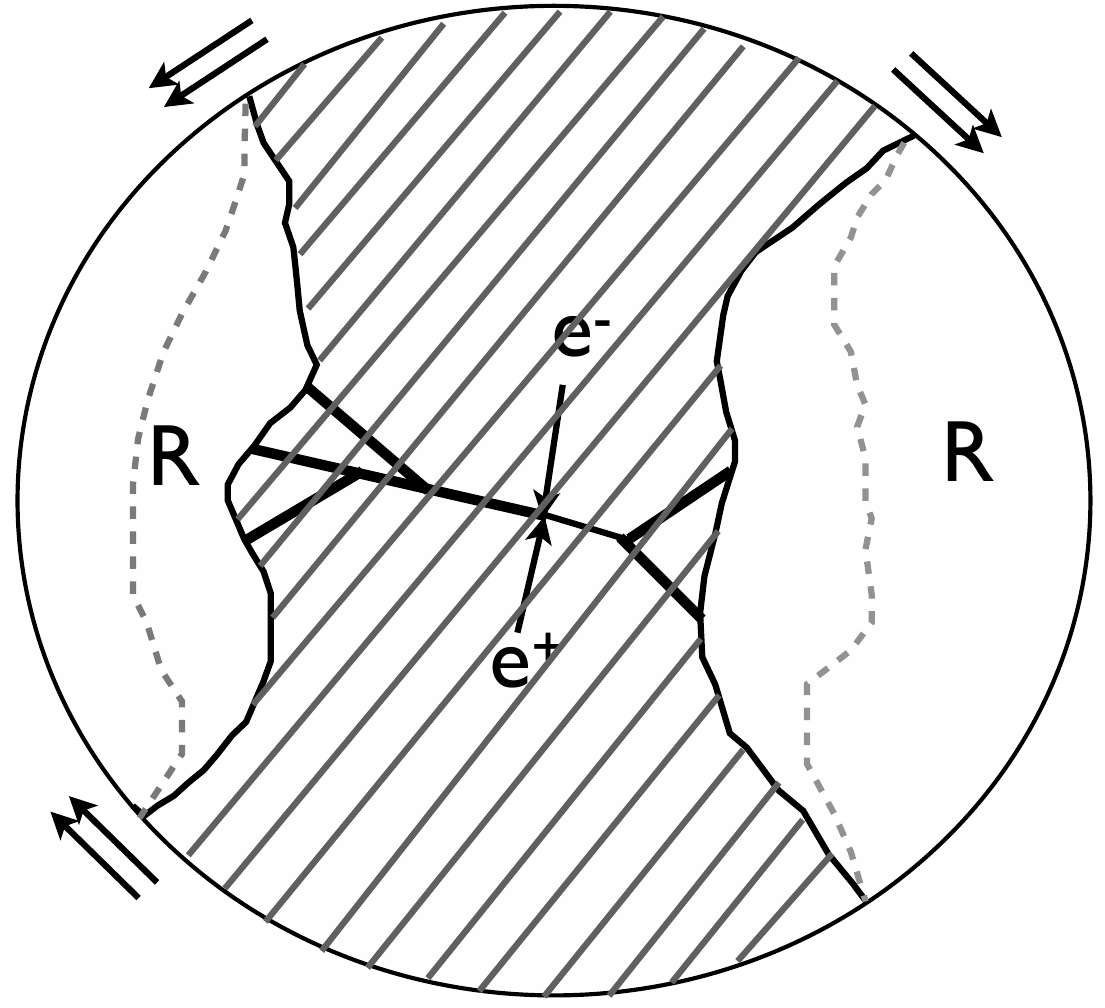}
\caption{Minimal example of a non-global observable:
the total cross-section to produce particles inside a given potato-shaped allowed region $R$, allowing only a small total energy $E_{\rm out}$ outside of it.  In the limit $E_{\rm out}\to 0$, large logarithms need to be resummed, which suppress the cross-section:
the effective excluded region grows as the veto suppresses near-boundary radiation.}
\label{fig:potato}
\end{figure}

For concreteness, let us describe an archetypical cross-section to which the formalism directly applies,
which exhibit purely non-global logarithms in a minimal way: {\it potato-shaped cross-sections},
shown in fig.~\ref{fig:potato}.
Given some fixed angular ``potato'' region $R$ (on a two-sphere detector surrounding a beam), the question
is what is the total-cross section
to produce particles inside this region, vetoing the energy outside of $R$ to be less than a small cutoff $E_{\rm out}$.

We believe that this typifies the essential complications of richer and phenomenologically interesting observables,
such as the hemisphere mass function  (see \cite{Kelley:2011ng}), which describes the probability of finding a small invariant mass
in one hemisphere centered around a jet.  Indeed this is essentially the limit where
one of the two potatoes in figure \ref{fig:potato} shrinks to a narrow cone. (Boosting the allowed hemisphere to a narrow cone,
this describes more generally the probability of finding isolated jets with most of their energy inside a given cone size.)
Characterizing the narrow jet by its invariant mass rather than its radius however departs from the considered class of cross-sections,
as a small invariant mass leads to further collinear logarithms that need to be resummed.

In general we expect that the radiation from the narrow jet will be factorized from the rest of the process and essentially ``global'',
so that inclusion of these effects will possible without major changes.\footnote{{\it Note added in v2}: This extension has now
been successfully achieved \cite{Becher:2015hka,Becher:2016mmh}.}
We leave this to the future, and in this paper focus on observables that do not focus on features at small angular scales,
so as to be dominated by the physics of soft, wide-angle radiation.

A mathematical definition of such a cross-section is
\be
 \mathcal{R}(Q,E_{\rm out}) = \sum_n \int d\Pi_n \big|A_{Q\to n}\big|^2
  \theta\left(E_{\rm out}-\sum_{i=1}^n \theta_{\bar {R}}(p_i)E_i\right) \label{observab}
\ee
where $Q$ describes the color-singlet initial state under consideration (for example, a virtual photon in $e^+e^-$ annihilation),
with invariant mass $Q$, $\theta$ is the step function, and $\theta_{\rm out}(p_i)$ projects onto the complement $\bar{R}$ of the allowed (``potato") region, thus putting a veto excluding radiation outside $R$.
Large logarithms, termed ``non-global'' because the final phase space is not globally integrated over, occur when the out-energy $E_{\rm out}$ is very small compared to $Q$.

Let us now explain how the renormalization-group equation (\ref{RG_for_sigmahard})
can be used to resum these large logarithms. 
The basic idea is to separate the hard and soft scales $Q$ and $E_{\rm out}$.
The veto is at a soft scale, so in the hard scale sector we include all radiation and ignore the veto, but weight radiation by a matrix $U(\theta)$ depending on the angle:
\be
 \sigma[U;Q]  \equiv \sum_n \int d\Pi_n \left[A_{n}^{a_1\cdots a_n}\right]^*
U^{a_1b_1}(\theta_1)\cdots
U^{a_nb_n}(\theta_n)  \left[A_{n}^{b_1\cdots b_n}\right]
\label{LL_sigma}
\ee
One could add any hard-scale vetoes to this, for example requiring that certain quantum numbers be present (e.g. charm)
or that a certain number of jets be present inside the potatoes according to some infrared-safe and collinear-safe jet definition.
In the case of the hemisphere mass function (used in the first arXiv 
version of this paper), for example, the invariant mass inside the ``heavy'' hemisphere would be set at this stage,
but no constraints are put yet on what happens in the light hemisphere.

The quantity (\ref{LL_sigma}) is of the form of the density matrix (\ref{density_matrix}).
It contains infrared divergences caused by the $U$ matrices, which as described exponentiate
and are to be renormalized at a scale $\mu$ (see eq.~(\ref{factorization})).
Concretely, in perturbation theory, $\sigma^{\ren}$ is polynomial in $U$'s and can be viewed
as a bookkeeping device encoding the orientations of outgoing partons.
In $e^+e^-$ annihilation to hadrons it would start with a dijet term
\be
 \sigma^{\ren}[U;Q,\mu] = C\int d^2\Omega_n {\rm Tr}\big[ U(n)U^\dagger(\bar{n}) \big] + O(g^2), \label{dijet_term}
\ee
where $n$ is a null vector integrated over a two-sphere, and $\bar{n}$ is pointing in the opposite direction.
The $U$ matrices associated with fermion jets are in the fundamental representation.

Choosing $\mu\sim Q$, the problem becomes single-scale and
$\sigma^{\rm ren}$ is given as a series in $\alphas(Q)$ which contains no large logarithms.
The idea is to use the RG equation (\ref{RG_for_sigmahard}) to run $\mu$ down
to the scale $E_{\rm out}$, where we deal with the infrared part of the measurement.
In the leading-log approximation, the $\theta(E_{\rm out}-\ldots)$ factor in the observable (\ref{observab})
simply removes all the out-radiation generated so far, so the IR measurement can be phrased in terms of an averaging
\be
 \mathcal{R}(Q,E_{\rm out}) = \langle \sigma^{\ren}[U;Q,\mu=E_{\rm out}]\rangle_{\rm IR}  \label{def_averaging}
\ee
where to leading order in the coupling\footnote{
Even though the right-hand-side is not a unitary matrix, the equation makes sense because the average of unitary matrices need not be
unitary.}
\be
 \l U_1^{a_1b_1} \cdots U_n^{a_nb_n} \r_{\rm IR} = \l U_1^{a_1b_1}\r_0 \cdots\l U_n^{a_nb_n}\r_0 + O(g^2) \quad\mbox{with}\quad \l U^{ab}(\theta_i) \r_0 =\delta^{ab}\theta_R(i)\,. \label{IR_measurement}
\ee
The step function allows real radiation inside the allowed region $R$, projecting to zero all the terms in $\sigma^{\rm ren}$
with $U$-matrices outside of it.

Note that the averaging procedure depends only on angles, since $\sigma[U]$ does not carry information about parton energies.
Operationally speaking, from the viewpoint of the soft physics, each $U$ matrix represents a hard parton and can thus be treated
as a Wilson line operator
(going from the origin in the matrix element to infinity, then back to the origin in the complex conjugate matrix element).
The details of the measurement on soft radiation, as defined by eq.~(\ref{observab}) or possible variations of it,
in general are encoded into $O(g^2)$ loop corrections to (\ref{IR_measurement}),
while hard physics including the possibility of 3-jet events (purely virtual for narrow jets)
are accounted for by the $O(g^2)$ term in eq.~(\ref{dijet_term}).
More precisely, the details of the IR measurement are encoded through Wilson line expectation values:
\be
 \l U_1^{a_1b_1} \cdots U_n^{a_nb_n} \r_{\rm IR} \equiv
 \sum_{n=0}^\infty\int d\Pi_n
 \big\l 0\big|U_1^{a_1}\cdots U_n^{a_n}\big|n\big\rangle\big\l n\big | U_1^{b_1}\cdots U_n^{b_n}\big|0\big\rangle
\times u^{\rm soft}(\{p_n\})  \label{IR_measurement1}
\ee
where now the sum runs over the soft partons in the final state (all hard partons having been replaced by the Wilson lines),
and $u^{\rm soft}$ accounts for that part of the measurement function $u(\{ p_n\})$ entering eq.~(\ref{finite_observable}) 
which has not yet been accounted for when defining $\sigma^{\rm ren}$.
At this stage, each Wilson line $U$ appearing inside a forbidden region should also be set to zero.
The Wilson lines extend to infinity along straight null lines, and the indices on them ``live" at the origin, where they meet
and are contracted into color-singlets according to the hard processes included in $\sigma^{\rm ren}$.

The IR measurement (\ref{IR_measurement}) is IR-finite but has ultraviolet divergences, which are to be renormalized
using the same scheme as the infrared divergences of $\sigma^{\rm ren}$, ensuring that the physical observable given by
eq.~(\ref{def_averaging}) is finite and scheme-independent.
The ultraviolet divergences include not only the usual ones present at the cusps, present in both the matrix element and its complex conjugate,
but also come from real radiation in the allowed region where partons can have arbitrary high energy in (\ref{IR_measurement1}).
The latter ultraviolet divergences nontrivially pair the amplitude and its conjugate.
Our proposal implies that the ultraviolet divergences of such defined
cusp Wilson line operators precisely match the infrared divergences of the color density matrix.

The reader may wonder why the excluded region is only projected out in the final step, in the IR in eq.~(\ref{IR_measurement1}),
rather than in the UV in eq.~(\ref{dijet_term}).
After all, why keep track of radiation in places that are not going to contribute in the end?
The answer is that $U$-matrices in forbidden regions can be dropped at any stage,
because the evolution equation only ever \emph{adds} $U$-matrices but never removes them.
However, by doing the projection in the IR, we make it possible for the evolution equation to be \emph{universal}:  the evolution
kernel $K$ is independent upon the shape of the exclusion region. This is a very useful and important property.

Comparison of this procedure with the leading-log prescriptions of refs.~\cite{Dasgupta:2001sh,Banfi:2002hw,Hatta:2013iba}
is discussed in section \ref{sec:review}.  In keeping with the logic of factorization, in this paper we will concentrate on the universal soft wide-angle
evolution $K$, and leave the UV and IR endpoint factors (the analogs of OPE coefficients) to future work.

A remarkable fact about $K$ is that it is also essentially the Balitsky-Fadin-Kuraev-Lipatov (BFKL) Hamiltonian, that is, the boost operator of the theory in the high-energy limit.
The same Hamiltonian $K$ thus simultaneously governs non-global logarithms and the Regge limit.
This was observed mathematically from the one-loop expressions in refs.~\cite{Banfi:2002hw,Weigert:2003mm}.
A general explanation has been given using a conformal transformation, which extends to higher loop orders \cite{Hatta:2008st}.
One thus anticipates the difference in QCD to be at most proportional to the $\beta$-function.

\begin{figure}
\centering
\be
\begin{array}{c@{\hspace{2cm}}c}
\def\svgwidth{6cm}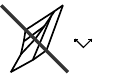&
\def\svgwidth{4cm}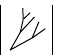\\
\hspace{-2cm}(a)&(b)\end{array}\nonumber
\ee
\caption{(a) Scattering in the Regge limit. The thin shock  is the Lorentz-contracted target.
(b) Branching of soft gluons.
To connect the pictures one `folds' along the target and sends it to infinity.
}\label{fig:bfkl}
\end{figure}

Since this correspondence will be technically useful it is helpful to include a rough explanation here.
High-energy forward scattering (for example the elastic $pp\to pp$ amplitude)
amounts to taking an instantaneous snapshot of a hadron's wavefunction, so pictorially it measures the amplitude for
a virtual shower to form inside the hadron and then recombine.  This is illustrated in fig.~\ref{fig:bfkl}(a).
This is also roughly what the density matrix $\sigma_{Q\to (\cdots)}[U]$ of a decaying virtual hadron measures.
Importantly, however, one measurement is instantaneous while the other takes place at infinity.
To relate them requires a conformal transformation as in ref.~\cite{Hatta:2008st}.

In this correspondence with the Regge limit, the color rotations in $\sigma[U]$ implement
the shockwave of the Balitsky-JIMWLK framework \cite{Balitsky:1995ub,JalilianMarian:1996xn,JalilianMarian:1997gr,Iancu:2001ad}.
Here the `shockwave' is inserted at infinity between the matrix element and its complex conjugate.
This was our original motivation for defining eq.~(\ref{density_matrix}). (Mathematically
similar considerations were used in refs.~\cite{Cornalba:2008qf,Cornalba:2009ax} to exploit
the conformal symmetry of the BFKL equation.)
Note that in the Regge context there is a natural symmetry between the projectile and target impact factors.
In the present context these correspond respectively
to the UV and IR measurements (\ref{LL_sigma}) and (\ref{IR_measurement}), and this symmetry is not obvious
(and broken by the running coupling).

The aim of this paper is to analyze the properties of the Hamiltonian $K$
and to calculate it explicitly to the next-to-leading order.  The lessons learned from this calculation
will then lead to an immediate proof of all-order exponentiation.  As a cross-check of the calculation
we will compare against results obtained in the context of Regge limit scattering.  

This paper is organized as follows.
In section \ref{sec:review} we review known facts regarding the exponentiation of infrared divergences and factorization of soft emissions.
We illustrate the formulas by giving the leading terms in perturbation theory for the various ingredients.
We also verify that the procedure around eq.~(\ref{IR_measurement}) reproduces the established resummation of non-global logarithms at leading-logarithm order.
In section \ref{sec:nlo} we perform the two-loop calculation. A key finding will be the possibility to express all terms in $K$ as finite integrals over
well-defined, finite and gauge-invariant building blocks.   The final result is recorded in subsection \ref{ssec:final_result}.
In section \ref{sec:bfkl} we compare our result for $K$ against the two-loop BFKL equation.  We will find
perfect agreement in conformal theories, with, as expected, a relatively compact correction term proportional to the $\beta$-function in QCD.
In section \ref{sec:higher_orders}, using the lessons learned from the two-loop calculation
we derive formal expressions for $K$ at three-loop and beyond, and demonstrate
exponentiation in general.  Conclusions are in section \ref{sec:conclusion}.
A technical appendix reports complete details of our evaluation of the real-virtual contributions at two-loop.

\noindent {\bf Note added in v2}:  Shortly after the first arXiv submission of this paper, the work \cite{Larkoski:2015zka}
discussed the resummation of non-global logarithms using the dressed gluon approximation (extending \cite{Forshaw:2008cq} and earlier work),
which is closely related to expanding in powers of $U$ the evolution in the present paper.
A closely related evolution equation for multi-parton Wilson lines operators (playing the role of our $U$ matrices)
is also obtained in \cite{Becher:2015hka,Becher:2016mmh}, who further discuss the factorization of collinear logarithms.

\section{Conventions and review}
\label{sec:review}

To set our conventions we now review the exponentiation of infrared divergences and give explicit formulas for the relevant objects at one loop.
We also discuss the resummation of non-global logarithms at leading-log.

\subsection{Infrared factorization}

As described in refs.~\cite{Sterman:2002qn,Aybat:2006mz,Feige:2014wja},
the exponentiation of infrared and collinear divergences is controlled by a soft anomalous dimension:
\be
 A_n = \mathcal{P} \exp\left[-\int_0^\mu \frac{d\lambda}{\lambda} \Gamma_n(\lambda,\alphas(\lambda),\eps)\right] \times H_n(\mu,\alphas(\mu),\eps)\,. \label{factorization_virtual}
\ee
For a gentle(r) introduction we refer to ref.~\cite{Sterman:2014nua}.
The infrared-renormalized amplitude $H_n$, also called the hard function,
is finite as $\eps\to 0$ (in this paper we use only ultraviolet-renormalized amplitudes).
The trade-off is that it depends on a factorization scale $\mu$:
\be
 \left[\mu\frac{\partial}{\partial\mu} + \beta(\alphas)\frac{\partial}{\partial\alphas}\right]H_n(\mu,\alphas(\mu),\eps) = \Gamma_n(\mu,\alphas(\mu),\eps) H_n(\mu,\alphas(\mu),\eps)\,. \label{RGforM}
\ee
It is important to note that since $\Gamma_n$ acts as a matrix in the space of color structures, the path-ordering symbol
cannot be omitted.  The fact that infrared divergences are controlled by a renormalization group equation, reflects, of course, the general
Wilsonian principle of decoupling between disparate length scales.  Indeed, eq.~(\ref{factorization_virtual}) can be obtained
by integrating eq.~(\ref{RGforM}) to the deep infrared where the $S$-matrix element $A_n$ is defined.

We work in $D=4{-}2\eps$ dimensions and the coupling constant depends on scale through
\be
 \mu\frac{\partial}{\partial\mu} \alphas(\mu,\eps) = \beta(\alphas,\eps), \qquad \beta(\alphas,\eps) = -2\eps\,\alphas + \beta(\alphas)\,.
\ee
At one-loop $\beta(\alphas)=-2\frac{\alphas^2}{4\pi}\betazero$ with $\betazero=\frac{11C_A-4\nF\TF-\nS\TS}{3}$ in a theory with $\nF$ flavors of Dirac fermions and $\nS$ complex scalars
(in QCD $\CA=3$ and $\TF=\frac12$).
The solution is then
\be
 \alphas(\mu) = \alphas(\mu_0)\frac{\mu^{-2\eps}}{\mu_0^{-2\eps}} \left[1+\frac{\betazero\alphas(\mu_0)}{2\pi} \frac{1-\mu^{-2\eps}/\mu_0^{-2\eps}}{2\eps}\right]^{-1}\,.
\ee
The integral in (\ref{factorization_virtual}) thus converges and produces the desired $1/\eps$ poles provided that $\eps$ is negative enough that the coupling vanishes in the infrared.

In the literature $\Gamma_n$ is often written as being $\eps$-independent, 
which defines minimal subtraction schemes.
We keep the more general notation since below we will also use a non-minimal scheme.
As long as $\Gamma_n$ remain finite as $\eps\to 0$, different schemes are related simply
by finite renormalizations of $H_n$.

In the soft limit, amplitudes with $m$ soft gluons factorize in a simple way \cite{Catani:1999ss,Feige:2014wja}
\be
 H_{n+m}^{\mu_1\cdots \mu_m,a_1\cdots a_m}(\mu,\alphas(\mu),\eps) \to
 \SS_m^{\mu_1\cdots \mu_m,a_1\cdots a_m}(k_1,\ldots, k_m;\mu,\alphas(\mu),\eps) \times H_n(\mu,\alphas(\mu),\eps)\,, \label{soft_factorization}
\ee
up to powers of $k$ provided that all of $\{k_1,{\ldots},k_m\}$ are softer than other scales in $M$.
We have only indicated the color and Lorentz indices of the soft gluons (to be contracted with polarization vectors $\epsilon_i^{\mu_i}$).
Since $H_n$ is the same as in eq.~(\ref{factorization_virtual}), this formula states that soft gluons can be `tacked onto' an amplitude without recomputing it.

Similarly to $\Gamma_n$, the soft currents $\SS_m$ are matrices in the space of color structures
of the hard partons. According to eq.~(\ref{RGforM}) they are finite and have factorization scale dependence
\ba
 \left[\mu\frac{\partial}{\partial \mu} + \beta(\alphas,\eps)\frac{\partial}{\partial\alphas}\right]\SS_m(\mu,\alphas(\mu),\eps) &=&
 \Gamma_{n{+}m}(\mu,\alphas(\mu),\eps) \SS_m(\mu,\alphas(\mu),\eps)
 \nl && - \SS_m(\mu,\alphas(\mu),\eps)\Gamma_n(\mu,\alphas(\mu),\eps)\,. \label{RGforSS}
\ea

Our main proposal is that the color density matrix admits a similar factorization,
\be
 \sigma[U] = \mathcal{P}\exp\left[-\int_0^\mu \frac{d\lambda}{\lambda} \GG(\lambda,\alphas(\lambda))\right] \sigma^{\ren}[U;\mu,\alphas(\mu),\eps]\,,
 \label{factorization}
\ee
where $\sigma^{\ren}[U]$ is finite and obeys the RG equation (\ref{RG_for_sigmahard}) quoted in the introduction.
Furthermore $K$ is independent of the measurement function $u(\{p_i\})$ appearing in the definition of $\sigma[U]$.
These will be shown in section \ref{sec:higher_orders}
to be essentially combinatorial consequences of the known factorization properties of soft gluons,
as stated in eqs.~(\ref{factorization_virtual}) and (\ref{soft_factorization}).

\subsection{Leading-order expressions}

The tree-level emission of one soft gluon is controlled by Weinberg's well-known soft current:
\be
 \SS^{\mu,a}_1 = g \sum_i  R_i^a S_i^\mu(k_1) + O(g^3)\quad\mbox{where}\quad S_i^\mu(k_1) = \frac{\b_i^\mu}{\b_i{\cdot}k_1}\,, \label{LOsoft}
\ee
where $R_i^a$ is the operator which inserts a color generator on leg $i$ and
$\beta_i^\mu=(1,\vec{v}_i)^\mu$ is a null vector proportional to $p_i$.
These obey $[R_j^a,R_k^b]=if^{abc}\delta_{jk}R_j^c$ and our normalizations are such that $\Tr[T^aT^b]=\frac12\delta^{ab}$ and $\Tr[1]=N_c$.
The soft anomalous dimension at one-loop is \cite{Sterman:2002qn}\footnote{The linear dependence on $\log\mu$
ensures the correct soft-collinear double poles upon integrating in eq.~(\ref{factorization_virtual}).}
\be
 \Gamma_n^{(1)} = -2\sum_{i\neq j} R_i^aR_j^a \log \frac{-2p_i{\cdot}p_j}{\mu^2} + \sum_i \gamma_i^{(1)}\, \label{LOgamma}
\ee
where the collinear anomalous dimensions are $\gamma_g^{(1)}=-\betazero$ for gluons and $\gamma_q^{(1)}=-3C_F$ for quarks.
We will loop-expand using the uniform notation:
$\Gamma_n=\sum_{\ell=1}^\infty \left(\frac{\alphas c_\Gamma}{4\pi}\right)^\ell\Gamma_n^{(\ell)}$
with
\be
c_\Gamma=\frac{\Gamma(1+\eps)\Gamma(1-\eps)^2}{\Gamma(1-2\eps)(4\pi)^{-\eps}}\,.
\ee

\begin{figure}
\centering
\be
\begin{array}{c@{\hspace{2cm}}c}
\def\svgwidth{6cm}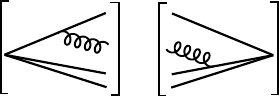&
\def\svgwidth{6cm}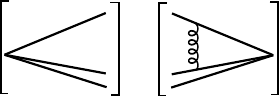\\
(a)&(b)\end{array}\nonumber
\ee
\caption{One-loop evolution of the density matrix.
(a) Real emission of one soft gluon. This adds one $U$-matrix (cf. eq.~(\ref{Gamma0_real})).
(b) Virtual correction.}
\label{fig:oneloop_real}
\end{figure}

To find $\GG$ at one loop it suffices to compute
$\sigma[U]$ to that accuracy and compare the divergence with eq.~(\ref{factorization}).
The real emission contribution to $\sigma[U]$ has an infrared divergence when an additional gluon
is emitted at a wide angle, as shown in fig.~\ref{fig:oneloop_real}(a).
It is given as the square of the soft current (\ref{LOsoft}):
\be
 \GG^{(1)}\big|_{\rm real} = \sum_{i,j} \int \frac{d^2\Omega_0}{4\pi} K^{(1)}_{ij;0} U^{aa'}_{0}(L_i^{a} R_j^{a'}{+}R_i^{a'}L_j^{a})\label{Gamma0_real}
\ee
where $K^{(1)}_{ij;0}$ is the (infrared) pole from the energy integral,
\be
 K^{(1)}_{ij;0} = 2\int_0^\mu daa^{1-2\eps}\,S_i^\mu(a\b_0) S_j^\mu(a\b_0)\Big|_{1/[2\epsilon]} = \frac{\z{i}{j}}{\z{0}{i}\z{0}{j}} \label{int_for_H1}
\ee
with $\z{i}{j}=\frac{-\b_i{\cdot}\b_j}{2}=\frac{1-\cos\theta_{ij}}{2}$.
Here the sums run over the $U$-matrices present in $\sigma[U]$ (which at finite order in perturbation theory is a polynomial)
and we use the abbreviation $U_k^{ab}=U^{ab}(\beta_k)$.
The operator $R_i^a$, as in eq.~(\ref{LOsoft}), is a color rotation in the amplitude. Specifically, here,
$R_i^a$ is the operator which replaces $U_i$ with $U_iT^a$. Similarly $L_i^a$, representing the color charge in the complex conjugate
amplitude, replaces $U_i$ with $T^aU_i$. These obey:
\be
 [R_j^a,R_k^b]=if^{abc}\delta_{jk}R_j^c\,,\qquad  [L_j^a,L_k^b]=-if^{abc}\delta_{jk}L_j^c\,,\qquad [R,L]=0\,.\nonumber
\ee

The virtual corrections (fig.~\ref{fig:oneloop_real}(b)) generate products of the type $LL$ and $RR$ with no extra $U$.
An important constraint is that $\sigma^{\ren}[U^{ab}=\delta^{ab}]$ must be evolution-invariant,
since this correspond to the total cross-section which is finite by the KLN theorem.
That is $\GG$ must vanish when $U$ is the identity field.
This unambiguously determines the $LL$ and $RR$ terms.
Using the identities
\be
 U_i^{aa'}R_i^{a'} = L_i^a, \qquad L_i^a U_i^{aa'}=R_i^{a'}\,,  \label{collinear_identities}
\ee
which in particular yield $L_i=R_i$ when $U_i^{ab}=\delta^{ab}$, the (unique) solution is easily seen to be\footnote{
A term $\sum_{i,j}f_{ij}(L_i^aL_j^a-R_i^aR_j^a)$ would also satisfy the KLN theorem
and preserve the reality of $\sigma$ provided that its coefficient is imaginary.
The imaginary part of the explicit expression (\ref{LOgamma}) however shows that $f_{ij}\propto i\pi$ is constant
and thus cancels out using color conservation in the case that all partons are outgoing.}
\be
\framebox{$\displaystyle
 \GG^{(1)} = \sum_{i,j} \int \frac{d^2\Omega_0}{4\pi} K^{(1)}_{ij;0}
 \left(U^{aa'}_{0}(L_i^{a} R_j^{a'}{+}R_i^{a'}L_j^{a})-L_i^{a}L_j^{a}-R_i^aR_j^a\right) \label{Gamman_0}\,.
 $}
\ee
This gives the complete scale dependence of the density matrix $\sigma^{\ren}[U]$, including non-planar effects
(and therefore, by the expected factorization, any non-global logarithm at leading-log).

We review a few known facts about this equation.
\begin{itemize}
\item Taking the 't Hooft planar limit $N_c\to\infty$ with $\lambda=g^2N_c$ fixed,
eq.~(\ref{Gamman_0}) becomes for the dipole $U_{ij}=\frac{1}{N_c}\Tr\big[U_i U_j^\dagger\big]$:
\be
 \GG\,U_{12} = \frac{g^2}{16\pi^2} \int\frac{d^2\Omega_0}{4\pi} \frac{2\z{1}{2}}{\z{0}{1}\z{0}{2}} \left(2C_F U_{12}-\frac{2}{N_c}\Tr\big[T^a U_1 T^{a'}U_2^\dagger\big]U_0^{aa'}\right) + O(g^4)\,.
\ee
Using simple color identities this reduces to a closed nonlinear equation:\footnote{We have used: $U_0^{aa'}T^{a'}=U_0^\dagger T^a U_0$ and $\Tr\big[T^a X T^a Y\big]=\frac12\Tr[X]\Tr[Y]-\frac{1}{2N_c}\Tr[X Y]$.}
\be
 \GG\, U_{12}= \frac{\lambda}{8\pi^2} \int\frac{d^2\Omega_0}{4\pi}\frac{\z{1}{2}}{\z{0}{1}\z{0}{2}} \left( U_{12}-U_{10}U_{02}\right) + O(\lambda^2,1/N_c)\,.
\label{LOplanar}
\ee
This is the Banfi-Marchesini-Smye (BMS) equation governing non-global logarithms in the planar limit \cite{Banfi:2002hw}.\footnote{
In addition, compared with ref.~\cite{Schwartz:2014wha} which deals with the hemisphere function,
one needs to set $U_{12}^{\rm here}=\theta_R'(1)\theta_R'(2)U_{12}^{\rm there}$; the step-functions
factors are stable under evolution. At leading-log collinear divergences exponentiate independently
so the $R$ term in $\theta_R'$ in eq.~(\ref{IR_measurement}) does not interfere with the non-global part.}
Let us be more precise.  As stated in the introduction, the \emph{functional} RG equation
(\ref{RG_for_sigmahard}) is to be integrated from the UV to the IR, starting from e.g. the dijet initial condition $\sigma[U]=U_{n\bar n}$ (\ref{dijet_term}).
In the IR one performs the average (\ref{IR_measurement}).  In the planar limit, the averaging reduces
to evaluating the functional at one point, $\sigma[U_{ij}=\theta_R'(i)\theta_R'(j)]$ corresponding to the step function in the infrared,
so the procedure is equivalent to evolving the \emph{argument} of the functional,
e.g. the function $U_{ij}$, from the IR to the UV.  This is precisely the procedure of \cite{Banfi:2002hw}.
\item
Away from the planar limit, eq.~(\ref{Gamman_0}) coincides with the generalization of the BMS formula derived in ref.~\cite{Weigert:2003mm}.
Again, as in the footnote for the BMS case, the two forms differ only by multiplication of the $U$ matrices by step functions, which commute with the evolution.
The averaging procedure (\ref{IR_measurement}), performed in the infrared, is as in refs.~\cite{Weigert:2003mm}.
\item The work of ref.~\cite{Weigert:2003mm} was at least partially motivated by analogy with
equations describing the Regge limit.  Using the substitution given below in eq.~(\ref{mapping1})
(namely, $\alpha_{ij}\to (x_{i}{-}x_{j})_\perp^2$ and $\int \frac{d^2\Omega_i}{4\pi} \to \frac{d^2x_\perp}{\pi}$),
eq.~(\ref{Gamman_0}) is indeed recognized as the Balitsky-JIMWLK equation \cite{Balitsky:1995ub,JalilianMarian:1996xn,JalilianMarian:1997gr,Iancu:2001ad}.
\item
The double-sum notation in eq.~(\ref{Gamman_0}) is most natural in a perturbative context where $\sigma[U]$ is a polynomial in the $U$'s.
Since the evolution increases the number of $U$'s,
for solution it can be better to view $\GG$ as a functional differential operator acting on $\sigma[U]$. This is achieved by the
following simple substitutions (done after normal-ordering all $L,R$'s to the right of $U$'s) \cite{Weigert:2000gi,Blaizot:2002np}:
\be
 \sum_i \mapsto \int d^2\Omega_i,\quad
 L_i^a \mapsto  (T^aU(\beta_i))\frac{\delta}{\delta U(\beta_i)},\quad
 R_i^a \mapsto (U(\beta_i)T^a)\frac{\delta}{\delta U(\beta_i)}\,. \label{integro_differential}
\ee
These $L_i^a$ and $R_i^a$ obey the same commutation relations as those defined previously,
and in fact after substitution into eq.~(\ref{Gamman_0}) one finds the same action on any polynomial $\sigma[U]$.
This reveals eq.~(\ref{Gamman_0}) as a functional second-order differential equation of the Fokker-Planck type,
whose solution can be importance-sampled via lattice Monte-Carlo techniques
 \cite{Weigert:2000gi,Blaizot:2002np}.  For studies of $1/N_c$ effects in the context of non-global logarithms,
 see \cite{Hatta:2013iba,Khelifa-Kerfa:2015mma}.
\item
Also well-studied is the weak-field regime where all matrices are close to the identity.
Following ref.~\cite{Caron-Huot:2013fea} and references therein, one writes $U_j = e^{igT^aW^a_j}$ and
expand (\ref{integro_differential}) in powers of $W$. This can be streamlined using
the Baker-Campbell-Hausdorff formula, which gives
\be
 ig L_j^a,ig R_j^a=  \frac{\delta}{\delta W_j^a} \pm \frac{g}2 f^{abx}W_j^x\frac{\delta}{\delta W_j^b}+\frac{g^2}{12} f^{aex}f^{eby}W_j^xW_j^y \frac{\delta}{\delta W_j^b} -\frac{g^4}{720}\big(\cdots\big)+\ldots \label{linearization}
\ee
Plugging this into the one-loop Hamiltonian yields after a small bit of algebra
(only the first two terms contribute)
\be
 \GG^{(1)} = f^{aa'c}f^{bb'c}\int \frac{d^2\Omega_{0}}{4\pi} \frac{d^2\Omega_1d^2d\Omega_2\,\z{1}{2}}{\z{0}{1}\z{0}{2}}
 (W_1^{a'}{-}W_0^{a'})(W_2^{b'}{-}W_0^{b'}) \frac{\delta^2}{\delta W_1^a\delta W_2^b}
 \label{linearized_bfkl}
\ee
up to nonlinear terms of the form $\delta\GG\sim g^4W^4\frac{\delta^2}{\delta W^2}$.
This is one form of the one-loop BFKL equation and its (`BJKP') multi-Reggeon generalization \cite{Kuraev:1977fs,Balitsky:1978ic},
valid for color-singlet states (see ref.~\cite{Caron-Huot:2013fea} and references therein). 
It acts on functionals $\sigma[W]$ where $W^a$ is identified as the Reggeized gluon field.  This identification will
play a useful role later in this paper.
\item
Finally, we did not prove in this subsection that divergences do exponentiate according to eq.~(\ref{factorization}). We simply read off
the exponent from a one-loop fixed-order calculation. Proofs to leading-logarithm accuracy are in refs.~\cite{Banfi:2002hw,Weigert:2003mm}
and an all-order demonstration is given in section \ref{sec:higher_orders}.
\end{itemize}

\section{Evolution equation to next-to-leading order}
\label{sec:nlo}

We now present a calculation of $\GG$ to the next-to-leading order,
by matching two-loop infrared divergences in $\sigma[U]$ against eq.~(\ref{factorization}).
The computation will be phrased exclusively in terms of convergent integrals over building blocks with a clear physical interpretation
(renormalized soft currents), which will shed light on the exponentiation mechanism.
We perform the computation in a general gauge theory, although at intermediate steps
we only write formulas for color-adjoint matter.
The reader not interested in the technical details
can skip directly to the final result in subsection \ref{ssec:final_result}.

\subsection{Building blocks: soft currents}

\begin{figure}
\centering
\be
\begin{array}{c@{\hspace{1.5cm}}c}
\raisebox{-0.8cm}{\def\svgwidth{3.5cm}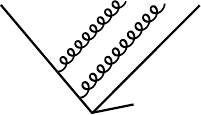}
\,\,+\,\,
\raisebox{-0.8cm}{\def\svgwidth{3.5cm}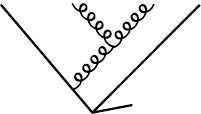}
& 
\raisebox{-0.8cm}{\def\svgwidth{3.5cm}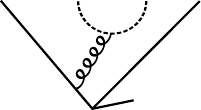}
\vspace{2mm}\\ 
(a)&(b)\end{array}
\nonumber\ee
\caption{Building block for next-to-leading order computation: amplitude for two soft particles.
Solid lines are eikonal Wilson lines. (a) Two soft gluons. The non-abelian part of the first graph
gives a connected contribution. (b) Two soft fermions or scalars.
} \label{fig:first_building_block}
\end{figure}

The first building block is the tree-level amplitude for emitting two soft gluons.
It can be written naturally as a sum of disconnected and connected parts:
\be
 \SS^{\mu\nu,ab}(k_1,k_2) =  g^2\sum_{i,j}  R_i^a R_j^b \,S_i^\mu(k_1) S_j^\nu(k_2) + g^2 \sum_i if^{abc}R_i^c S_{i}^{\mu\nu}(k_1,k_2) + O(g^4)\,, \label{double_real_tree}
\ee
with $S_i^\mu(k_1)=\frac{\b_i^\mu}{\b_i{\cdot}k_1}$ the one-gluon soft current as above. The connected part
\be
S_{i}^{\mu\nu}(k_1,k_2) = \frac{1}{2\b_i{\cdot}(k_1{+}k_2)}
 \left[ \frac{\b_i^\mu \b_i^\nu}{\b_i{\cdot}k_1}-\frac{\b_i^\mu \b_i^\nu}{\b_i{\cdot}k_2} + \frac{\delta^{\mu\nu}\b_i{\cdot}(k_2{-}k_1)+2(\b_i^\mu k_1^\nu-k_2^\mu \b_i^\nu)}{k_1{\cdot}k_2}\right] \label{Soft2}
\ee
follows directly from the Feynman graphs shown in fig.~\ref{fig:first_building_block}(a) \cite{Catani:1999ss}.
Here and below, to optimize the notation, all color generators are implicitly symmetrized: $R_i^a R_j^b \to \frac12 \{ R_i^a,R_j^b\}$, which is relevant
when $i=j$. This notational convention (borrowed from ref.~\cite{Gardi:2013ita}) ensures that the connected part is proportional to $f^{abc}$.

To familiarize ourselves with the notation we review the transverseness check: $k_{1\mu} \SS^{\mu\nu}=0$.
For the individual $S_i$ one finds
\be
 k_{1\mu} S_i^\mu(k_1) =1\,, \qquad
 k_{1\mu} S_i^{\mu\nu}(k_1,k_2) = \frac12\left( \frac{k_1^\nu}{k_1{\cdot}k_2} - \frac{\b_i^\nu}{\b_i{\cdot}k_2}\right)\,. \label{Ward}
\ee
We need to use color conservation in the form of the identity $(\sum_{i=1}^n R_i^a)H_n=0$.  Since
this holds when $\sum_i R_i$ is inserted to the right of an operator product, the implicit
symmetrization in eq.~(\ref{double_real_tree}) produces commutators.  For example the divergence of the first sum is
\be
k_{1\mu} \sum_{i,j}  R_i^a R_j^b \,S_i^\mu(k_1) S_j^\nu(k_2) \equiv
\sum_{i,j} \frac12\{R_i^a,R_j^b\}\,S_j^\nu(k_2) = \sum_{i} \frac12[R_i^a,R_i^b]S_i^\nu(k_2)
=\frac{i f^{abc}}{2}\sum_i R_i^c S_i^\nu(k_2)\,.\nonumber
\ee
This is easily seen to cancel the second term in the parenthesis in (\ref{Ward}), up to a $\b_i$-independent term which itself cancels
due to $\sum_i R_i^c=0$, thus proving transverseness.

Pairs of soft fermions or soft scalars can also be emitted (fig.~\ref{fig:first_building_block}(b)).
For notational simplicity we carry out all intermediate steps
in a theory with $n_{\rm Weyl}^{\adj}$ color-adjoint Weyl fermions and $n_s^{\adj}$ real adjoint scalars
(the final result will be trivial to generalize). Then:
\be
 \SS^{f\bar f}(k_1,k_2) = ig^2\sum_i f^{abc} S_i^{f\bar f}(k_1,k_2),\quad
 \SS^{ss}(k_1,k_2) = ig^2\sum_i f^{abc} S_i^{ss}(k_1,k_2)
\ee
with
\be
 S_i^{f\bar{f}}(k_1,k_2) = \frac{[2|\b_i|1\r}{2\b_i{\cdot}(k_1{+}k_2)k_1{\cdot}k_2},\quad
 S_i^{ss}(k_1,k_2) =\frac{\b_i{\cdot}(k_2{-}k_1)}{2\b_i{\cdot}(k_1{+}k_2)k_1{\cdot}k_2}\,.
\ee

\begin{figure}
\centering
\be
\raisebox{-0.8cm}{\def\svgwidth{3.5cm}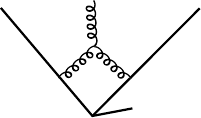}
\,\,+\,\,
\raisebox{-0.8cm}{\def\svgwidth{3.5cm}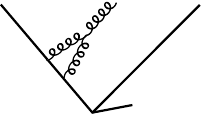}
\,\,+\,\,
\raisebox{-0.8cm}{\def\svgwidth{3.5cm}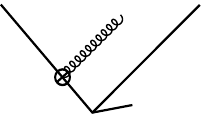}
\nonumber\ee
\caption{Second building block: one-loop soft current.
} \label{fig:second_building_block}
\end{figure}

The second building block is the next-to-leading order soft gluon amplitude $\SS_1^{(1)}$.
Representative graphs are shown in fig.~\ref{fig:second_building_block}, however the result has been computed a long time ago
by taking the soft limit of a five-parton amplitude and comparing with the four-point amplitude \cite{Bern:1998sc,Bern:1999ry,Catani:2000pi}.
These references give the factorization of the amplitude $A_n$ and contain $1/\eps^2$ and $1/\eps$ infrared divergences.
To convert to our infrared-finite soft current, which gives the factorization of the hard function (\ref{soft_factorization}),
we need the account for the renormalization factor in eq.~(\ref{factorization_virtual}), which at one-loop
simply removes the pole terms (and nothing else).
Up to $O(\eps)$ terms, this gives:
\be
 \SS_1^{(1)\mu a}(k_1) = g^3\sum_i R_i^a S_i^\mu + ig^3\sum_{i,j} f^{abc} R_i^b R_j^c S_{ij}^{(1)\mu}(k_1) \label{SS1_virtual_colors}
\ee
with
\ba
S_i^{(1)\mu}(k)&=&-\frac{\pi^2\CA}{6}\frac{\b_i^\mu}{\b_i{\cdot}k}\,,\nl 
S_{ij}^{(1)\mu}(k) &=& \frac12 \left(\frac{\b_j^\mu}{\b_j{\cdot}k}{-}\frac{\b_i^\mu}{\b_i{\cdot}k}\right) \log^2\left(\frac{(-2\b_i{\cdot}k-i0)(-2\b_j{\cdot}k-i0)}{\mu^2(-2\b_i{\cdot}\b_j-i0)}\right)\,.\nonumber
\ea
This is also transverse.  Note that the constant term has been extracted from $S_{ij}$ and put into $S_i$ using color conservation.
All coupling constants are evaluated at the scale $\mu$,
and the $\mu$-dependence agrees precisely with the renormalization group equation (\ref{RGforSS}).\footnote{
One might be surprised that the one-loop $\beta$-function does not explicitly appear in the soft current,
given that the tree-level coupling $\sim g$ should produce some scale dependence.
This gets canceled because the one-loop gluon collinear anomalous dimension
happens to equal precisely $-b_0$ (see eq.~(\ref{LOgamma})).
}

This is all we will need!
From eq.~(\ref{factorization}), the next-to-leading order kernel is given as the divergent part (coefficient of $1/(4\eps)$) of the following combination:
\be
 \GG^{(2)}\sigma^{(0)\ren}= \left(\sigma^{(2)}-\frac{\GG^{(1)}}{2\eps}\sigma^{(1)\ren}-\frac12\left(\frac{\GG^{(1)}}{2\eps}\right)^2
  \sigma^{(0)\ren}
  +\frac{\betazero\GG^{(1)}}{4\eps^2} \sigma^{(0)\ren}\right)_{1/[4\eps]}. \label{Gamma2fromsigma}
\ee
We will now see that this can be expressed in terms of the soft currents given above.
 
\subsection{Double-real emission}

\def\doublereal{{\rm double\,real}}
\def\singlereal{{\rm single\,real}}

We begin with the terms in the NLO kernel which involve two wide-angle soft partons, and thus generate two additional $U$ factors.
The double-real contribution to $\sigma^{(2)}$ is by definition (suppressing color indices)
\be
\sigma^{(2)}\big|_{\doublereal} \equiv \int\frac{d^{2{-}2\eps}\Omega_0}{\pi(2\pi)^{-2\eps}c_\Gamma} \frac{d^{2{-}2\eps}\Omega_{0'}}{\pi (2\pi)^{-2\eps}c_\Gamma}U_0 U_{0'}
\!\!\!\!\int\limits_{0<a<b<\infty} \!\!\!\!a^{1{-}2\eps}da\, b^{1{-}2\eps}db \,|A_{n{+}2}|^2(a\b_0,b\b_{0'})\,. \label{sigma2real}
\ee
(The integrals have compact support due to the momentum-conserving $\delta$-function in $A$,
and we do not show a factor $d\Pi_n u(\{p_i\})$ associated with the underlying hard event.)
The trick to evaluate (\ref{Gamma2fromsigma}) is to find compatible integral representations for $\GG^{(1)}$ and $\sigma^{(1)\ren}$.
For $\GG^{(1)}$ we already have eq.~(\ref{int_for_H1}) and subtracting it from $\sigma^{(1)}$ leaves simply
\be
 \sigma^{(1)\ren}\big|_{\rm real} = \int \frac{d^{2{-}2\eps}\Omega_0}{\pi (2\pi)^{-2\eps}c_\Gamma}U_0 \int_\mu^\infty a^{1{-}2\eps}da \,|A_{n{+}1}|^2(a\b_0)\,. \label{sigma_one_real}
\ee
The essential point here is that the matrix element factorizes in the soft region, $|A_{n{+}1}|^2(a\b_0)\to |\SS_{n} A_{n}|^2$,
so that subtracting $\frac{\GG^{(1)}}{2\eps}\sigma^{(0)}$ is equivalent to removing the integration region $a<\mu$ (to all orders in $\eps$).
Invoking factorization similarly, eq.~(\ref{Gamma2fromsigma}) can be re-written as:
\ba
\GG^{(2)}\big|_{\doublereal} &\propto&
\int\limits_{0<a<b<\infty} F(a,b)
-\int\limits_{\substack{0<a<\mu\\ \mu<b<\infty}} F(a,b)\big|_{a\ll b} -\int\limits_{0<a<b<\mu} F(a,b)\big|_{a\ll b}
\nl &=& \int\limits_{\substack{0<a<\mu\\ a<b<\infty}}\left(F(a,b)-F(a,b)\big|_{a\ll b}\right) + \int\limits_{\mu<a<b<\infty}F(a,b)\,. \label{double_real_regions}
\ea
Here $F(a,b)$ denotes the integrand in (\ref{sigma2real}). This formula is also exact in $\eps$. One can see
that the second integral is finite and the first integral has no subdivergences.
(Except from collinear regions, which are dealt with in the next subsection.)
After scaling $a\to a b$ in the first integral to extract the pole, one thus just get:
\be
\framebox{$\displaystyle
 \GG^{(2)}\big|_{\doublereal} = -\int\frac{d\Omega_0}{\pi} \frac{d\Omega_{0'}}{\pi}U_0U_{0'}
 \int_0^1 ada \left(\SS(a\b_0,\b_{0'})\SS(a\b_0,\b_{0'})^* - \big|_{1/a^2}\right)\,.$} \label{double_real_master}
\ee
This is the desired formula, which expresses the double-real contribution to $\GG^{(2)}$ as a convergent integral
over tree-level soft currents.  The integrand measures the extent to which two soft emissions are not independent from each other.

\begin{figure}
\centering
\be
\begin{array}{c@{\hspace{0.8cm}}c@{\hspace{0.8cm}}c}
\def\svgwidth{4.5cm}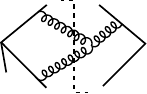&
\def\svgwidth{4.9cm}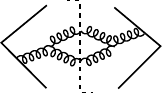&
\def\svgwidth{4.5cm}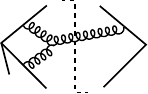
\end{array}\nonumber\ee
\caption{Sample two-loop diagrams included in eqs.~(\ref{double_real_colors}) and (\ref{colors_virtual}).
}\label{fig:twoloop}
\end{figure}

Using the explicit expressions (\ref{double_real_tree}) the formula yields two nontrivial color structures
\ba
\GG^{(2)}\big|_{\doublereal} &=&
\phantom{+}\sum_{i,j,k} \int \frac{d^2\Omega_0}{4\pi}\frac{d^2\Omega_{0'}}{4\pi} K^{(2)}_{ijk;00'}
if^{abc}\left( L_i^{a'}L_j^{b'}U_0^{a'a}U_{0'}^{b'b} R_k^c-R_i^{a'}R_j^{b'}U_0^{aa'}U_{0'}^{bb'} L_k^c\right)
\nl
&& + 
\sum_{i,j} \int \frac{d^2\Omega_0}{4\pi}\frac{d^2\Omega_{0'}}{4\pi} K^{(2)}_{ij;00'}
f^{abc}f^{a'b'c'}U_{0}^{bb'}U_{0'}^{cc'}\left(L_i^aR_j^{a'}+R_i^{a'}L_j^a\right) \label{double_real_colors}
\ea
(shown in fig.~\ref{fig:twoloop}). These multiply angular functions
\be\begin{aligned}
\!K^{(2)}_{ijk;00'} \!\!=& {-}8 \!\int_{0}^1 \!\!\!ada
\left(S_i^\mu(a\b_0) S_j^\nu(\b_{0'})S_k^{\mu\nu}(a\b_0,\b_{0'}) +S_i^\mu(\b_0) S_j^\nu(a\b_{0'})S_k^{\mu\nu}(\b_0,a\b_{0'})- \big|_{1/a^2}\right)
\\
K^{(2)}_{ij;00'} \!\!=& {-}4\!\int_0^1 \!\!\!ada
\left(S_i^{\mu\nu}(a\b_0,\b_{0'})S_j^{\mu\nu}(a\b_0,\b_{0'})
+ S_i^{\mu\nu}(\b_0,a\b_{0'})S_j^{\mu\nu}(\b_0,a\b_{0'})
+{\rm matter}- \big|_{1/a^2}\right)\!.
\end{aligned}\nonumber \ee
We note the absence of a fully disconnected (Abelian) color structure: since its squared amplitude is proportional to $1/a^2$ it disappeared before integration.
The matter contribution in the last parenthesis is, in full: $n^{\adj}_{\rm Weyl} (S_i^{f\bar f}S_j^{\bar f f}+S_i^{\bar f f}S_j^{f\bar f}) + n_s^{\adj} S_i^{ss}S_j^{ss}$.
The $a$-integrals are elementary (and convergent!) and after some straightforward algebra starting from (\ref{Soft2}) yield
\begin{subequations}
\ba
K^{(2)}_{ijk;00'} &=& \frac{1}{\z{0}{i}\z{0'}{j}}
\left[ \frac{\z{i}{j}}{\z{0}{0'}} + \frac{\z{i}{k}\z{j}{k}}{\z{0}{k}\z{0'}{k}}-
\frac{\z{0}{j}\z{i}{k}}{\z{0}{k}\z{0}{0'}}- \frac{\z{j}{k}\z{0'}{i}}{\z{0}{0'}\z{0'}{k}}\right]
\log\frac{\zsq{0}{k}}{\zsq{0'}{k}}\,,\\
 K^{(2)}_{ij;00'} &=& \frac{\z{i}{j}}{\z{0}{i}\z{0}{0'}\z{0'}{j}}\left[ 1+ \frac{\z{0}{0'}\z{i}{j}}{\z{0}{i}\z{0'}{j}-\z{0'}{i}\z{0}{j}}\right]\log\frac{\z{0}{i}\z{0'}{j}}{\z{0'}{i}\z{0}{j}}
 +\frac{\big(n_{\rm Weyl}^{\adj}{-}4\big)}{\z{0}{0'}}\frac{\z{i}{j}\log \frac{\z{0}{i}\z{0'}{j}}{\z{0'}{i}\z{0}{j}}}{\z{0}{i}\z{0'}{j}-\z{0'}{i}\z{0}{j}}
  \nl &&
 + \frac{\big(n_s^{\adj}-2n_{\rm Weyl}^{\adj}+2\big)}{2\zsq{0}{0'}}
 \left[\frac{\z{0}{i}\z{0'}{j}+\z{0'}{i}\z{0}{j}}{\z{0}{i}\z{0'}{j}-\z{0'}{i}\z{0}{j}}\log\frac{\z{0}{i}\z{0'}{j}}{\z{0'}{i}\z{0}{j}} -2\right]\,.
\ea
\label{realcorrections}
\end{subequations}
\!\!For $K^{(2)}_{ij;00'}$ we have used symmetry in $(i\leftrightarrow j)$ to simplify.
The expression is especially compact in $\mathcal{N}=4$ SYM (the first term).  The second and third terms represent,
respectively, an (adjoint) $\mathcal{N}=1$ chiral multiplet and a scalar. The rational structures are such that all potential
divergences associated with the $\beta_i,\beta_j$ and $\beta_k$ regions cancel.
There remains a divergence as $\beta_{0}\to\beta_{0'}$, proportional to the gluon collinear anomalous dimension
$\gamma_g^{(1)}\propto\betazero$,
which will be canceled shortly.

At this point we could stop: using simple and not-so-simple physical considerations discussed below
one could determine the full result using only what we have so far.
We find it instructive, however, to continue with the explicit computation.

\subsection{Single-real emission}
\label{ssec:singlereal}

\def\Splitsq#1#2{\big|{\rm Split}_{#1}({#2})\big|^2}

We now turn to terms with one radiated parton; since these contain only one $U$ field at a wide angle,
these will combine with and cancel the collinear divergences in eq.~(\ref{realcorrections}).

According to the factorization formula (\ref{factorization}), the correction to the two-loop kernel
will come from the infrared divergence of the one-loop single-real emission, minus iteration of leading-order effects
of the form $K^{(1)}\big|_{\rm real} K^{(1)}\big|_{\rm virtual}$.
One may thus anticipate that the subtraction will convert the emission amplitude
to its finite renormalized version $\SS$ defined in eq.~(\ref{soft_factorization}).
This would be the case if the virtual part of $K^{(1)}$ precisely matched the usual soft anomalous dimension $\Gamma_n$.
This is not exactly correct, due to
the different ways in which they treat collinear regions,
however assuming that $K\big|_{\rm virtual}\simeq \Gamma_n$ will provide useful intuition.
Let us thus first ignore the difference and begin by writing the single-real contribution in terms of the hard function (\ref{factorization_virtual}):
\be
\sigma_{\singlereal} \equiv
 \int  \frac{d^{2{-}2\eps}\Omega_0}{4\pi (2\pi)^{-2\eps}c_\Gamma}U_0 \int_0^\infty da a^{1{-}2\eps}\,
\left|\mathcal{P} e^{-\int_0^\mu \frac{d\lambda}{\lambda} \Gamma_{n{+}1}} H_{n+1}(a\b_0;\mu)\right|^2\,.
\ee
In the soft region we can replace the amplitude by a soft current times $H_n$.
It is useful to run the soft current to its natural scale, $\mu=a$,
using eq.~(\ref{RGforSS}). The energy integral then becomes, formally to all orders in perturbation theory,
\be\begin{aligned}
&\phantom{+}\int_0^\mu da a^{1{-}2\eps}
 \left|\mathcal{P} e^{-\int_0^a \frac{d\lambda}{\lambda} \Gamma_{n{+}1}} \bar{\SS}_1(a\b_0;a)\right|^2
 \left|\mathcal{P}e^{-\int_a^\mu \frac{d\lambda}{\lambda} \Gamma_n} H_n\right|^2 \\&\hspace{0cm} +
 \int_\mu^\infty da a^{1{-}2\eps}\left| \mathcal{P} e^{-\int_0^\mu \frac{d\lambda}{\lambda} \Gamma_{n{+}1}} H_{n{+}1}(a\b_0)\right|^2\,.
 \label{one_loop_master_pre}
\end{aligned}\ee
Comparing against the factorization formula (\ref{factorization}),
and pretending that $\Gamma_n\simeq K\big|_{\rm virtual}$,
we see that the second integral represents a finite correction to
the finite coefficient $\sigma^{\ren}$. The $a$-integrand on the first line, on the other hand, is nicely identified as
the following shift to the exponent:
\be
 \framebox{$\displaystyle
 \GG
 \big|_{\singlereal} 
 =-\int \frac{d^2\Omega_0}{\pi} U_0 \, \big| \bar{\SS}_1(\mu\b_0;\mu)\big|^2
 $}\,. \label{one_loop_master}
\ee
This gives the single-real contribution to $K$, formally to all loop orders. This generalizes in the simplest conceivable way the leading-order result: one simply evaluates the loop-corrected soft current with energy set equal to the renormalization scale $\mu$.
Because the soft current is used at its natural scale, the series for (\ref{one_loop_master}) contains no large logarithms and
the $\betazero$ term in eq.~(\ref{Gamma2fromsigma}) is automatically accounted for.

The `bar' on $\bar{\SS}_1$ is now to account for the discrepancy between $\Gamma_n$ and $K^{(1)}\big|_{\rm virtual}$.
Indeed, at the amplitude level, $\Gamma_n$ contains collinear divergences, whereas for the angularly weighted cross-sections
that we are interested in, the collinear divergences cancel between real and virtual corrections.
Thus to precisely define the relation between $\bar{\SS}_1$ and $\SS_1$,
we first need to precisely define the subtraction which will make the double-real term (\ref{realcorrections}) well-defined in its
collinear limits.

We do so by adding and subtracting the integral of using universal splitting functions.
First, we make the one-parton emission eq.~(\ref{sigma_one_real}) finite by adding an integral of splitting function (for each external particle),
minus their soft limits:
\be\begin{aligned}
 \sigma^{(1)\ren}\big|_{\rm real} &\!\!\to\!\!
 \int \!\!\frac{d^{2{-}2\eps}\Omega_0}{4\pi (2\pi)^{-2\eps}c_\Gamma}\int_0^\infty \!\!a^{1{-}2\eps}da
 \left(4U_0 \big|A_{n{+}1}(a\b_0)\big|^2- \sum_i \theta(k_i{-}a)\Splitsq{i}{a\beta_0;k_i}\big|A_n\big|^2
 \right.\\&\hspace{2.5cm}\left. -\big|_{1/a^2}\theta(\mu-a)
 +\sum_i\theta(\mu{-}a)\Splitsq{i}{(k_i-a)\beta_0;k_i}\big|A_n\big|^2\big|_{1/a^2}\right). \label{real_subtraction}
\end{aligned}\ee
The next-to-last term removes the $a\to 0$ limit of everything to its left, and the momenta $k_1,\ldots k_n$ of the
parent amplitude $A_n$ are left untouched.  The formula differs from eq.~(\ref{sigma_one_real})
only by the splitting functions, which are added in such a way as to introduce no new soft divergences. 
The function ${\rm Split}_i(p_1;k_i)$ (representing the amplitude for particle $i$ to split into two with momenta $a\b_0$ and $k_i-a\b_0$,
symmetrized between the two),
is required to have the same integrand-level collinear singularity as $|A_{n{+}1}|^2$ when $\beta_0\parallel k_i$,
ensuring convergence.  This is guaranteed to exist by the factorization of amplitudes in collinear limits,
and an explicit expression is given in (\ref{split}).

We stress that the subtraction is not written as an integral over the phase space of $(n+1)$ particles with the original total momentum, since $a$ and $\b_0$ do not enter $A_n$. 
For subtraction schemes which fully account for momentum conservation, see for example dipole subtraction \cite{Catani:1996vz}.
The simpler scheme here will suffice for our purposes.

Since we're just shuffling terms between $\sigma^{(1)\ren}\big|_{\rm virtual}$
and $\sigma^{(1)\ren}\big|_{\rm real}$ but do not wish to modify their sum (which is already fully determined by $K^{(1)}$ and
soft- and collinear-finite),
one needs to make a compensating subtraction in the virtual contribution:
$\sigma^{(1)\ren}\big|_{\rm virtual} =2{\rm Re}[\bar{H}_n^{(1)}H_n^{(0)}]$, where $\bar{H}$ subtracts what was just added:
\be\begin{aligned}
 \bar{H}_n^{(1)} &= A_n^{(1)} - \int_0^\mu \frac{da}{a} \frac{a^{-2\eps}}{\mu^{-2\eps}}
 \int  \frac{d^{2{-}2\eps}\Omega_0}{4\pi (2\pi)^{-2\eps}c_\Gamma}\left( \sum_{i\neq j}\frac{\z{i}{j}R_i^aR_j^a}{\z{0}{i}\z{0}{j}}
+\sum_i a^2\Splitsq{i}{a\b_0;k_i}\big|_{a\to 0}\right)A_n^{(0)}
\\&\hspace{0.8cm}
+\frac12\sum_i\int_0^{k_i} \frac{da\,a^{1-2\eps}}{\mu^{-2\eps}}
\int  \frac{d^{2{-}2\eps}\Omega_0}{4\pi (2\pi)^{-2\eps}c_\Gamma}
\Splitsq{i}{a\b_0;k_i}A_n^{(0)}+\sum_i\frac{i\pi C_i}{\eps}A_n^{(0)}\,.  \label{primedscheme}
\end{aligned}\ee
This defines an alternative \emph{hard function}, which, just like $H_n$, is finite as $\eps\to 0$,
as verified in appendix.
(The imaginary part, which cancels in the cross-section, has been added by hands so that this also holds for the imaginary part.)
Indeed, one can see that the integral of $R_i^a R_j^a$ term basically gives $\GG^{(1)}$, up to a constant.
The barred hard functions can thus viewed as simply the hard functions in a different scheme, see specifically eq.~(\ref{Xexpression}).

The subtractions now made well-defined at the level of $\sigma^{(1)\ren}\big|_{\rm real}$ can now be applied to the double-real part of the kernel
as done in the preceding subsection.  It is important that we did not introduce any new soft divergences,
so that all the iterative subtractions of soft limits still work through, for example, in the double-real term in eq.~(\ref{sigma2real})
one simply subtracts $\big(\Splitsq{g}{a\b_{0},b\b_{0'}}+\sum_i\Splitsq{i}{a\b_{0},k_i}\big)\big|A_{n{+}1}(b\b_{0'})\big|^2$
from the integrand. In the $\Gamma^{(1)}$ terms one subtracts only the soft limit.
In this way all two-particle-collinear divergences are removed from the preceding subsection,
at the only cost of adding a piece to eq.~(\ref{double_real_master}):
\be
\framebox{$\displaystyle
\GG^{(2)} \big|_{\substack{\doublereal\\ {\rm linear\, in\, } U}} =
\int\frac{d\Omega_0}{\pi}\frac{d\Omega_{0'}}{\pi}U_{0} 
\int_0^1 \frac{ada}{1+a} \left(\frac{1}{2}\Splitsq{0}{a\b_{0'},(1+a)\b_0}\,\big|\SS^{(0)}(\b_{0})\big|^2 
 - \big|_{1/a^2}\right)\,.
 $}
\label{double_real_master2}
\ee
The argument of the splitting function is such that most energetic particle of the pair has energy $\mu$ (which we scaled out).
This removes precisely the collinear divergence
of the explicit formula in eq.~(\ref{realcorrections}) (at the integrand level and to all orders in $\eps$, to leading power in small angle).

In summary, the total double-real plus single-real kernel is given as the sum of eqs.~(\ref{realcorrections}) and (\ref{double_real_master2}),
which is collinear-safe, plus eq.~(\ref{one_loop_master}) where the
soft current in the barred scheme is defined by (\ref{primedscheme}) and given explicitly in eq.~(\ref{Soneprime}).
These ingredients are finite and only their four-dimensional limits contribute to $K^{(2)}$.
The evaluation is conceptually straightforward and detailed in appendix \ref{app:virtual}.

\subsubsection*{Simple ansatz for single-virtual terms}

It turns out that the result could have been anticipated using (not so trivial) physical considerations,
so here we concentrate on explaining these considerations.
The least obvious consideration is gluon Reggeization, or, more broadly, the connection with BFKL.
As mentioned in introduction, $\GG$ is the BFKL Hamiltonian in disguise (up to $\beta$-function terms).
Interactions between Reggeized gluons are constrained by physical principles such as Hermiticity of the boost operator
and signature conservation $U^{ab}\to U^{ba}$, which are not self-evident from the perspective of non-global logarithms.

We consider the weak-field regime where $U=e^{igT^aW^a}$ with the `Reggeized gluon' field $W$ small.
Linearizing the Hamiltonian yields Reggeon-number conserving terms given in eq.~(\ref{linearized_bfkl}),
as well as $2\to 4$ transitions (between states with with different powers of $W$) at order $g^4$, and so on.
Hermiticity of the boost operator (with respect to the specific inner product given by the scattering amplitude of left- and right- Wilson lines)
then predicts $4\to 2$ transitions at the same order, whose existence is indeed well-known \cite{Gribov:1984tu}. These are the terms which close the so-called Pomeron loop.
Now when reverting to the current power-counting which treats $(U-1)$ as $O(1)$, instead of $O(g)$, these $4\to 2$ transition
become a three-loop effect (see ref.~\cite{Caron-Huot:2013fea} and references therein).  Signature forbids $3\to 2$ transitions.
Hence the remarkable statement that $\GG$ must be triangular at one- and two-loop \cite{Caron-Huot:2013fea}:
\be
 \left[ \int d^2\Omega_0 W_0^a \frac{\delta}{\delta W_0^a}, \GG^{(L)} \right]\geq 0 \qquad (L=1,2). \label{reggeization}
\ee
Mathematically, this formula (just with the $L=1$ case) can be seen as equivalent to gluon Reggeization,
since it ensures that sectors with different powers of $W$'s can be diagonalized independently at one loop.
One then expects the Reggeized gluon ($W$ field) to provide
a good degree of freedom upon which to organize the perturbative spectrum of $\GG$ to any order
(as usually happens in degenerate perturbation theory after degeneracies are lifted at one-loop).

\begin{figure}
\def\spc{\hspace{0.3cm}}
\centering
\be
\raisebox{-0.8cm}{\def\svgwidth{3.2cm}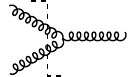}
\spc-\spc
\raisebox{-0.8cm}{\def\svgwidth{3.2cm}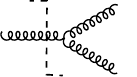}
\spc-\spc
\raisebox{-0.8cm}{\def\svgwidth{3.2cm}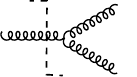}
\spc+\spc
\raisebox{-0.8cm}{\def\svgwidth{2.1cm}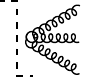}
\nonumber\ee
\caption{A combination of real and virtual color structures allowed by Reggeization.
}\label{fig:colorstructures}
\end{figure}

For our immediate purposes, eq.~(\ref{reggeization}) constrains two-loop color structures.
One easily sees that no double-real color structure satisfies it by itself:
for example, using (\ref{linearization}), the first line of eq.~(\ref{double_real_colors})
linearizes to give terms which replace three Reggeons $W_iW_jW_k$ by a single one $W_0$.
Cancelling this term uniquely fixes the range-three part of the single-real contribution (to the form in eq.~(\ref{ansatz_angular_function1})).
From other terms one constrains the double-virtual and range-two kernels.
In this way, using in addition that double-real terms are signature-even, we find that the
two-loop Hamiltonian can be parametrized by at most three angular functions:
\ba
\GG^{(2)} &=& 
\phantom{+}  \sum_{i,j,k} \int \frac{d^2\Omega_0}{4\pi}\frac{d^2\Omega_{0'}}{4\pi}K^{(2)}_{ijk;00'}
if^{abc}\left(\begin{array}{l}\phantom{+}
 (L_i^{a'}U_{0}^{a'a}{-}R_i^{a})(L_j^{b'}U_{{0'}}^{b'b}{-}R_j^{b})R_k^{c}
\\ -(L_i^a{-}U_{0}^{aa'}R_i^{a'})(L_j^{b}{-}U_{{0'}}^{bb'}R_j^{b'})L_k^{c}
\end{array}\right)
\nl
&&+ \sum_{i,j}\int \frac{d^2\Omega_0}{4\pi}\frac{d^2\Omega_{0'}}{4\pi}
 K^{(2)}_{ij;00'}
\left( f^{abc}f^{a'b'c'}U_{0}^{bb'}U_{0'}^{cc'}-\frac12C_A(U_{0}^{aa'} {+}U_{0'}^{aa'})\right) (L_i^aR_j^{a'}+R_i^{a'}L_j^a)
\nl
 &&
 +\sum_{i,j}\int \frac{d^2\Omega_0}{4\pi} K^{(2)}_{ij;0} \left(U^{aa'}_{0}(L_i^{a} R_j^{a'}{+}R_i^{a'}L_j^{a})-L_i^{a}L_j^{a}-R_i^aR_j^a\right)\,.
 \label{generalform}
\ea
The first one is shown in fig.~\ref{fig:colorstructures}.
Since $K^{(2)}_{ijk;00'}$ and $K^{(2)}_{ij;00'}$ have already been determined from double-real emissions,
effectively eq.~(\ref{reggeization}) predicts all virtual corrections, up to a term proportional to the leading-order structure (the last line).
The physical interpretation is that gluon Reggeization entails nontrivial real-virtual connections, which was indeed the original observation \cite{Kuraev:1977fs,Balitsky:1978ic}.

In appendix \ref{app:virtual} the prediction (\ref{generalform}) is compared with the direct evaluation of single-real terms
eqs.~(\ref{one_loop_master}), (\ref{double_real_master2}).  It turns out that there is a subtle loophole in the above argument:
in the non-global log context, $U^{ab}\to U^{ba}$ is not symmetry but only need to send the Hamiltonian to its complex conjugate.
Thus signature is not conserved.  The ansatz fails, by a single signature-odd term:
\be
 \GG^{(2)} \supset 2\pi i\sum_{i,j,k}if^{abc}\int \frac{d^2\Omega_0}{4\pi}
\left(\frac{\z{j}{k}}{\z{0}{j}\z{0}{k}}-\frac{\z{i}{k}}{\z{0}{i}\z{0}{k}}\right)\log\frac{\z{i}{j}}{\z{0}{i}\z{0}{j}}
 \left(L_i^aL_j^b U_0^{cc'}R_k^{c'}{+}L_k^{c'}U_0^{c'c}R_i^aR_j^b\right).
\label{imaginary_term}
\ee
The origin of this term is simple: the imaginary part
of the one-loop soft current (\ref{SS1_virtual_colors}). Its physical significance will be discussed shortly.
The explicit computation in appendix also yields the yet-undetermined signature-even function:
\be
 K^{(2)}_{ij;0} = 
 \frac{\z{i}{j}}{\z{0}{i}\z{0}{j}}\left[ \gammaK^{(2)}+\betazero\log \frac{\z{i}{j}}{4\z{0}{i}\z{0}{j}}\right]\,,\qquad
 \gammaK^{(2)}= -\frac{\pi^2\CA}{3}+\frac{64\CA}{9}-\frac{20\nF\TF}{9}-\frac{8\nS\TS}{9}\,.\label{H2_bare_version}
\ee
This contains only the two-loop cusp anomalous dimension \cite{Korchemsky:1987wg} (we use a normalization such that normalization $\gammaK^{(1)}=1$)
and the one-loop $\beta$-function.\footnote{
The appendix uses the so-called dimensional reduction scheme.
In conventional dimensional regularization (CDR), more commonly used in QCD,
a simple coupling redefinition \cite{Altarelli:1980fi} gives: $\frac{64\CA}{9}\to\frac{67\CA}{9}$.}
Physically, the $\betazero$ term is fully dictated by the collinear anomaly discussed subsection \ref{ssec:lorentz},
while the cusp anomalous dimension term can be checked to provide the correct Sudakov double logarithms in the limit of
a narrow jet cone.

How to explain the real-virtual connections (\ref{generalform}) from the perspective of non-global logarithms?
Perhaps one could use the Feynman tree theorem \cite{Feynman:1963ax}.
This is a way of computing loops by putting one (or more) propagator per loop on-shell.
Indeed, putting a gluon on-shell in fig.~\ref{fig:second_building_block} one can recognize diagrams of fig.~\ref{fig:first_building_block},
so at least schematically this seems to work.
The tree theorem was streamlined and generalized to higher loops, with at least partial success, in refs.~\cite{Bierenbaum:2010cy,CaronHuot:2010zt}; it would be interesting to see its implications in detail here.

\subsection{Double-virtual terms}

Let us now make sure that the ansatz (\ref{generalform}) is not missing any virtual corrections.
\emph{A priori} these could involve two color structures:
\be
 i f^{abc}\sum_{i,j,k}\big(h_{ijk}R_i^aR_j^bR_k^c-h_{ijk}^*R_i^aR_j^bR_k^c\big) + \sum_{i,j}\big(h_{ij}R_i^aR_j^a+h_{ij}^*L_i^aL_j^a\big)\,.
\ee
The coefficients are constrained by the KLN theorem:
at each loop order we can impose that $K$ vanishes exactly in $\epsilon$ when all Wilson lines are set to $U^{ab}=\delta^{ab}$.
(One might in principle consider schemes which impose only the weaker condition that $K$  be proportional to $\epsilon$ in this limit,
but it is always possible to impose the KLN condition exactly, as we did at one-loop (\ref{Gamman_0}).)
The other constraint is Lorentz invariance. Unfortunately, without adding signature symmetry, this does not fix them uniquely. (Below in eq.~(\ref{imaginary_term}) we give
an example of a signature-odd function satisfying both constraints.)
We resort to explicit computation.

The two-loop soft anomalous dimension is known to take the `dipole' form \cite{Aybat:2006mz,Becher:2009cu,Gardi:2009qi}
\be
\Gamma_n^{(2)} = -2\gamma_{\rm K}^{(2)}\sum_{i,j} R_i^aR_j^a \log\left(\frac{-2p_i{\cdot}p_j-i0}{\mu^2}\right)
+ \mbox{collinear terms}\,. \label{dipole_formula}
\ee
This gives the divergence of the amplitude after subtracting the square of $\Gamma_n^{(1)}$.
Since we are instead subtracting $\GG^{(1)}$, we need again to switch to the collinear-subtracted barred scheme (\ref{Xexpression}):
\be
 \bar{\Gamma}_n^{(2)} = \Gamma_n^{(2)}+[\bar{X}^{(1)},\Gamma_n^{(1)}] -2\betazero \bar{X}^{(1)} + \mbox{collinear terms}\,.
 \label{Gamma2_prime}
\ee
We omit `collinear terms' which depend on only one leg at a time,
since these are trivial to fix using the KLN theorem.
Concentrating on the terms which have nontrivial color structures and which are not so easily fixed,
the calculation of eq.~(\ref{Gamma2_prime}) is rather straightforward
and detailed in appendix \ref{app:virtual}. The outcome confirms that
no additional terms besides (\ref{imaginary_term}) need to be added to the Ansatz (\ref{generalform}).

We can now interpret this term (\ref{imaginary_term}).
First we observe that it can be mostly removed by a finite scheme transformation. Namely, if we set
\be
\sigma^{\ren}[U]^{\MSbar'}=\sigma^{\ren}[U]^{\MSbar}
 - \frac{\alphas}{4\pi}\sum_{i,j} i\pi\log(\z{i}{j})(L_i^aL_j^a-R_i^aR_j^a)\,\sigma^{\ren}[U]^{\MSbar} +O(\alphas^2,\eps)\,, \label{MSbar_primed}
\ee
where the $\MSbar$ density matrix is the minimally subtracted one we have been working with so far,
then the two-loop Hamiltonian in $\MSbar'$ gets shifted by a commutator with $\GG^{(1)}$ and a $\beta$-function term.
It is easy to check that the commutator term precisely cancels (\ref{imaginary_term}). The $\beta$-function term
then replaces it by
\be
 (\ref{imaginary_term}) \mapsto 2\pi i b_0 \sum_{i,j} \log(\z{i}{j})\big(L_i^aL_j^a-R_i^aR_j^a\big)\,. \label{unconstrained_function}
\ee
This combination is Lorentz-invariant in an interesting way: under a rescaling of $\b_i$, the $j$-sum become telescopic
and simplifies to $(C_i-C_i)=0$. This also satisfies the KLN theorem, being zero when $L=R$.
The existence of this structure is the only reason we needed
to use the explicit formula (\ref{dipole_formula}) to get the virtual corrections, otherwise the KLN theorem and Lorentz invariance would have sufficed.
It violates the triangular structure (\ref{reggeization}) but since it is proportional to the $\beta$-function
this does not contradict the BFKL-based argument leading to it.
Contrary to the imaginary part at one-loop, which canceled out telescopically in the case of color-singlet
initial state that we consider (as noted above eq.~(\ref{Gamman_0}),
the above does not cancel because each term has a different angular dependence.

The contribution (\ref{unconstrained_function}) has a simple and suggestive physical interpretation:
effectively it replaces the spacelike couplings in the one-loop evolution, by timelike counterparts:
\be
 g(\mu^2)R^a \to g(-\mu^2-i0)R^a,\qquad  g(\mu^2)L^a \to g(-\mu^2+i0)L^a\,. \label{timelike_coupling}
\ee
With hindsight, had we used timelike couplings in the one-loop evolution, we would never have had
to write down eqs.~(\ref{imaginary_term}), (\ref{MSbar_primed}) nor (\ref{unconstrained_function}).
We will nonetheless continue to use the (more conventional) spacelike coupling.

\subsection{Lorentz invariance and (lack of) collinear anomaly}
\label{ssec:lorentz}

We have assembled all ingredients of the kernel, but we notice
that the angular functions are not Lorentz-covariant: the arguments
of the logarithms (\ref{realcorrections}) are not homogeneous in $\b_0,\b_{0'}$ (and thus depend
on the frame choice implicit in the normalization $\b_i^\mu=(1,\vec{v}_i)$).
This may seem surprising given that dimensional regularization preserves Lorentz invariance.  

The simple explanation is that we did \emph{not} write the one-loop evolution in a $D$-dimensional covariant form.
What \emph{would} constitute a Lorentz-invariant version is instead:
\be
\frac{d^{2{-}2\eps}\Omega_0}{4\pi(2\pi)^{-2\eps}} K^{(1)}_{ij;0}\mapsto
\frac{4^\eps d^{2{-}2\eps}\Omega_0}{4\pi(2\pi)^{-2\eps}} \left(\frac{\z{i}{j}}{\z{0}{i}\z{0}{j}}\right)^{1{-}\eps}= 
\frac{d^{2{-}2\eps}\Omega_0}{4\pi(2\pi)^{-2\eps}} \left( K^{(1)}_{ij;0} +2\eps \delta^{(1)}_{ij;0} +\ldots\right)\,,
\label{scheme_eps}
\ee
which differs at order $\epsilon$ by the amount $\delta^{(1)}_{ij;0}=\frac{\z{i}{j}}{2\z{0}{i}\z{0}{j}}\log \frac{4\z{0}{i}\z{0}{j}}{\z{i}{j}}$.  
The integrand is now homogeneous in all of $\b_i,\b_j,\b_0$ and
one may check that under a Lorentz transformation the Jacobian factor precisely cancels the change in the parenthesis.
(The factor $4^\eps$ is for future convenience.)

An $O(\epsilon)$ shift to an anomalous dimension, as usual, is equivalent to a finite renormalization
of $\sigma^{\ren}$, e.g. a scheme transformation.
The density matrix in the `Lorentz' scheme (\ref{scheme_eps}) is related to the $\MSbar$ one used so far, or better the $\MSbar'$ scheme just defined in the preceding subsection, as:
\be\begin{aligned}
 \sigma^{\ren,\lorentz} &= \sigma^{\ren,\MSbar'} -\frac{\alphas}{4\pi}\sum_{i,j} \int \frac{d^2\Omega_0}{4\pi} \delta^{(1)}_{ij;0} \left(U^{aa'}_{0}(L_i^{a} R_j^{a'}{+}R_i^{a'}L_j^{a})-L_i^{a}L_j^{a}-R_i^aR_j^a\right)\sigma^{\ren,\MSbar'}\,.
\end{aligned}\label{finite_renorm}\ee
At two-loops, this shifts $K^{(2)}$ by a commutator $[\GG^{(1)}, \delta^{(1)}]$ as well as a $\beta$-function term.

This transformation is only well-defined because it contains both real and virtual terms: the middle integral in eq.~(\ref{scheme_eps}) would otherwise be un-regulated even for $\eps\neq0$.
This clash between Lorentz covariance and collinear divergences reflects the (now called) collinear anomaly of refs.~\cite{Becher:2009cu,Gardi:2009qi}. Here, the anomaly cancels between real and virtual terms and we obtain a kernel which is homogeneous in all $\beta_i$.

To make this fully manifest we must still manipulate algebraically the expression for the triple-sum and double-sum terms by using color conservation
to add terms independent of some of the $\b_i$, being careful with commutators
as below eq.~(\ref{Ward}).  Collecting these commutators is tedious but fortunately the task can be easily automated on a computer.
We find (as it should) that the color structures in eq.~(\ref{generalform}) are preserved under these operations (see also ref.~\cite{Kovner:2014xia}).
Thus using this freedom, parametrized by two functions $E$ and $F$, to change these coefficients without changing $K$ itself,
the two-loop evolution in the Lorentz-covariant scheme becomes
\ba
 K^{(2)\lorentz}_{ijk;00'} &=& K^{(2)\MSbar}_{ijk;00'}
 + 2\left(K^{(1)}_{ik;0}\delta^{(1)}_{jk;0'}+ \delta^{(1)}_{i0';0}K^{(1)}_{jk;0'} +\delta^{(1)}_{ik;0}K^{(1)}_{j0;0'}
 -(\delta^{(1)}\leftrightarrow K^{(1)})\right)
 \nl && + E_{ij;00'}+2F_{ik;00'}-2F_{jk;0'0}\nl
 K^{(2)\lorentz}_{ij;00'} &=& K^{(2)\MSbar}_{ij;00'} + \left(
 \delta^{(1)}_{j0;0'}K^{(1)}_{ij;0}{+}\delta^{(1)}_{i0';0}K^{(1)}_{ij;0'}
 -(\delta^{(1)}\leftrightarrow K^{(1)})\right){+}F_{ij;00'}{+}F_{ji;0'0} \nl
 K^{(2){\rm lorentz}}_{ij;0} &=& K^{(2)\MSbar}_{ij;0}
+2\betazero \delta^{(1)}_{ij;0} 
+\frac{\CA}{2}
\int \frac{d^2\Omega_{0'}}{4\pi}
\Big\{
\left(\big(K^{(1)}_{i0;0'}+K^{(1)}_{j0;0'}\big)\delta^{(1)}_{ij;0}-(K^{(1)}\leftrightarrow\delta^{(1)})\right)
\nl &&
\hspace{5cm} +E_{ij;00'}+F_{ij;0'0}+F_{ji;0'0}-(0\leftrightarrow0')\Big\}. \label{change_in_kernels}
\ea
The functions $E_{ij;0}$ and $F_{ij;0}$ are arbitrary, with $E_{ij;00'}=-E_{ji;0'0}$.
The formula (\ref{final_result}) below arises for
$E_{ij;00'}=\frac{\z{i}{j}}{\z{0}{i}\z{0}{0'}\z{0'}{j}}\log \frac{\z{0'}{i}\z{0'}{j}}{\z{0}{i}\z{0}{j}}$ and
$F_{ik;00'}=\frac{\z{i}{k}}{2\z{0}{i}\z{0}{0'}\z{0'}{k}}\log \frac{\z{i}{k}\z{0}{0'}}{\z{0}{k}\z{0'}{i}}$.
(With these choices all $K$'s become homogeneous in $\beta$'s. The integral on the last line vanishes.)

\subsection{Final result for the evolution equation}
\label{ssec:final_result}

\def\LU#1#2#3{L_{#2;#1}^{#3}}
\def\UR#1#2#3{R_{#2;#1}^{#3}}

We record our final result for the two-loop Hamiltonian in the `Lorentz' scheme (superscript $\ell$),
which combines eqs.~(\ref{generalform})--(\ref{H2_bare_version}) with the finite renormalizations (\ref{MSbar_primed}) and (\ref{finite_renorm}).
For convenience we repeat the color structures, switching to the integro-differential notation (\ref{integro_differential}):
\ba
\GG^{(2)} \!&=&\!\!\!\!
\phantom{+}  \int_{i,j,k}\! \int\!\frac{d^2\Omega_0}{4\pi}\frac{d^2\Omega_{0'}}{4\pi}K^{(2)\lorentz}_{ijk;00'}
if^{abc}\left(\LU{0}{i}{a} \LU{0'}{j}{b} R_k^c-\UR{0}{i}{a}\UR{0'}{j}{b}L_k^c\right)
\nl
&&\!\!\!\!+ \int_{i,j}\!\int\!\frac{d^2\Omega_0}{4\pi}\frac{d^2\Omega_{0'}}{4\pi}
 K^{(2)N=4,\lorentz}_{ij;00'}
\left( f^{abc}f^{a'b'c'}U_{0}^{bb'}U_{0'}^{cc'}-\frac{\CA}{2}(U_{0}^{aa'} {+}U_{0'}^{aa'})\right) (L_i^aR_j^{a'}{+}R_i^{a'}L_j^a)
\nl
 &&
 \!\!\!\!+\int_{i,j}\!\int\!\frac{d^2\Omega_0}{4\pi} \frac{\z{i}{j}}{\z{0}{i}\z{0}{j}} \gammaK^{(2)}\left(\UR{0}{i}{a}L_j^a+\LU{0}{i}{a}R_j^a\right) + \GG^{(2)N\neq4}
 \label{generalform1}.
\ea
Here $\z{i}{j}=\frac{-\beta_i{\cdot}\beta_j}{2}=\frac{1{-}\cos\theta_{ij}}{2}$, $\LU{0}{i}{a}\equiv(L_i^{a'}U_{0}^{a'a}-R_i^{a})$, $\UR{0}{i}{a}\equiv(U_{0}^{aa'}R_i^{a'}-L_i^a)$ and $\int_i=\int d^2\Omega_i$,
the color rotations $L$ and $R$ being differential operators
defined in eq.~(\ref{integro_differential}). All products of $L_i^a$'s and $R_i^a$'s are implicitly symmetrized and normal-ordered to the right of
$U_0$, $U_{0'}$.
The third term is simply the one-loop result (\ref{Gamman_0}) times
the cusp anomalous dimension (\ref{H2_bare_version}).  The angular functions are:
\ba
 \z{0}{i}\z{0'}{j}K^{(2)\lorentz}_{ijk;00'}&=&
\frac{\z{i}{j}}{\z{0}{0'}}
\log \frac{\z{0'}{i}\z{0'}{j}\zsq{0}{k}}{\z{0}{i}\z{0}{j}\zsq{0'}{k}}
+\frac{\z{i}{k}\z{j}{k}}{\z{0}{k}\z{0'}{k}}
\log \frac{\z{i}{k}\z{0'}{j}\z{0}{k}}{\z{j}{k}\z{0}{i}\z{0'}{k}}
+\frac{\z{0'}{i}\z{j}{k}}{\z{0}{0'}\z{0'}{k}}
\log \frac{\z{j}{k}\z{0}{i}\z{0}{0'}\z{0'}{k}}{\zsq{0}{k}\z{0'}{i}\z{0'}{j}}
\nl &&
-\frac{\z{i}{k}\z{0}{j}}{\z{0}{k}\z{0}{0'}}
\log \frac{\z{i}{k}\z{0'}{j}\z{0}{0'}\z{0}{k}}{\zsq{0'}{k}\z{0}{i}\z{0}{j}}
+\frac{\z{i}{k}\z{0'}{j}}{\z{0'}{k}\z{0}{0'}}\log\frac{\z{i}{k}\z{0}{0'}}{\z{0}{k}\z{0'}{i}}
-\frac{\z{0}{i}\z{j}{k}}{\z{0}{k}\z{0}{0'}}\log\frac{\z{j}{k}\z{0}{0'}}{\z{0'}{k}\z{0}{j}}
 \nl
K^{(2)N=4,\lorentz}_{ij;00'}&=&
 \frac{\z{i}{j}}{\z{0}{i}\z{0}{0'}\z{0}{'j}}
 \left(2\log \frac{\z{i}{j}\z{0}{0'}}{\z{0'}{i}\z{0}{j}}+
\left[1+ \frac{\z{i}{j}\z{0}{0'}}{\z{0}{i}\z{0'}{j}-\z{0'}{i}\z{0}{j}}\right]\log \frac{\z{0}{i}\z{0'}{j}}{\z{0'}{i}\z{0}{j}}\right). \label{final_result}
\ea
This is the complete result in $\mathcal{N}=4$ SYM. In a general gauge theory with $n_F$ flavors of Dirac fermions and $n_S$ complex scalars in the representation $R$,
there additional contributions from matter loops, also obtained in eq.~(\ref{realcorrections}).
Upon restoring group theory factors corresponding to representation $R$,
in accordance with the square of fig.~\ref{fig:first_building_block}(b), these can be written:
\ba
  \GG^{(2)N\neq4}
&=&\int_{i,j}\!\int \frac{d\Omega_0}{4\pi}\frac{d\Omega_{0'}}{4\pi} \frac{1}{\z{0}{0'}}
 \left[\frac{\z{i}{j}\log \frac{\z{0}{i}\z{0'}{j}}{\z{0'}{i}\z{0}{j}}}{\z{0}{i}\z{0'}{j}-\z{0'}{i}\z{0}{j}}\right](L_i^aR_j^{a'}+R_i^{a'}L_j^a)
\nl&&\hspace{0.2cm}\times \left\{2\nF\Tr_R\big[ T^a U_0 T^{a'}U_{0'}^\dagger\big]-4f^{abc}f^{a'b'c'}U_{0}^{bb'}U_{0'}^{cc'}
-(\nF\TR-2\CA)(U_{0}^{aa'} {+}U_{0'}^{aa'})\right\}
\nl&+&\int_{i,j}\!\int \frac{d\Omega_0}{4\pi}\frac{d\Omega_{0'}}{4\pi}\frac{1}{2\zsq{0}{0'}}
\left[\frac{\z{0}{i}\z{0'}{j}+\z{0'}{i}\z{0}{j}}{\z{0}{i}\z{0'}{j}-\z{0'}{i}\z{0}{j}}\log\frac{\z{0}{i}\z{0'}{j}}{\z{0'}{i}\z{0}{j}} -2\right]
(L_i^aR_j^{a'}+R_i^{a'}L_j^a)
\nl && \hspace{0.2cm}\times
\left\{\begin{array}{c}\displaystyle2(\nS-2\nF)\Tr_R\big[ T^a U_0 T^{a'}U_{0'}^\dagger\big]+2f^{abc}f^{a'b'c'}U_{0}^{bb'}U_{0'}^{cc'}\\
-((\nS-2\nF)\TR+C_A)(U_{0}^{aa'} {+}U_{0'}^{aa'})\end{array}\right\}
\nl &+&\int_{i,j} 2\pi i b_0 \log(\z{i}{j})\big(L_i^aL_j^a-R_i^aR_j^a\big).
\label{final_result_matter}\ea
All sums are individually Lorentz-invariant (invariant under rescalings of the individual $\b_i$).
The first term is the contribution of two chiral $\mathcal{N}=1$ multiplets (minus the four adjoints in $\mathcal{N}=4$ SYM) and the second term collects remaining scalars; $b_0=\frac13(11\CA-4\nF\TR-\nS\TR)$.

\section{Comparison with BFKL and conformal transformation}
\label{sec:bfkl}

As mentioned in the introduction, the same Hamiltonian $K$ governs the Regge limit.
Hence the reader familiar with the literature on the Regge limit, in particular the Balitsky-JIMWLK equation,
will have recognized several equations by this point.  Let us now discuss the connection in detail.

Physically, as sketched in the introduction, the connection originates from the existence of a conformal transformation
which interchanges the $x^+=0$ light-sheet and future (null) infinity.
This interchanges the target residing at $x^+=0$ with the color rotations in the definition (\ref{density_matrix}) of
$\sigma[U]$.
It is given explicitly as \cite{Hatta:2008st,Cornalba:2006xk,Hofman:2008ar}
\be
 y^+=\frac{1}{\mu^2 x^+}\,,\quad y_\perp=\frac{x_\perp}{\mu x^+},\quad y^-=x^- -\frac{x_\perp^2}{2x^+}
 \label{conformal}
\ee
where $\mu$ is a reference scale.
This maps the Minkowski metric $ds^2=-2dx^+dx^-{+}dx_\perp^2$ to a multiple of itself, as one may verify.
Points approaching the BFKL target, $x^+\to0$, are mapped to infinity along the null direction
$y^\mu\propto (\b^0,\b_\perp,\b_z) = (\frac{1+\mu^2x_\perp^2}{\mu},2x_\perp,\frac{1-\mu^2x_\perp^2}{\mu})$.
In this way the transverse plane of the BFKL problem is mapped stereographically onto the two-sphere at infinity
of the non-global log problem.
If it were the case that the conformal transformation (\ref{conformal}) preserved the Lagrangian, this map would predict
that $K$ should go into the BFKL Hamiltonian
upon substituting \cite{Hatta:2008st}\footnote{
Here we use a normalization $\b^0\neq 1$ which differs from that adopted elsewhere in the present paper and in ref.~\cite{Hatta:2008st}.
This has not effect in Lorentz-covariant expressions such as eqs.~(\ref{final_result})-(\ref{final_result_matter}),
but one should remember to include additional factors $\b^0=(1+\mu^2x_i^2)$ if using non-covariant formulas such as eq.~(\ref{realcorrections}).}:
\be
 \alpha_{ij}\equiv -\frac{\b_i{\cdot}\b_j}{2} \to (x_{i}-x_{j})_\perp^2\,,
 \qquad \int \frac{d^2\Omega_i}{4\pi} \to \frac{d^2x_\perp}{\pi}\,. \label{mapping1}
\ee
We now verify this equivalence directly,
beginning with the case of $\mathcal{N}=4$ SYM where conformal symmetry is unbroken.
After that we discuss the general case, where we anticipate a discrepancy proportional to the $\beta$-function.

\subsection{Comparison in $\mathcal{N}=4$ SYM}
\label{ssec:N4comp}

It is instructive to consider a special case: we act with $\GG^{(2)}$ on a dipole $U_{12}={\rm Tr}[U_1U_2^\dagger]$.
The form (\ref{final_result}) is particularly convenient for this since $K^{(2)\lorentz}_{ijk;00'}$ vanishes when $i=k$ or $j=k$. The only terms in the first line are thus $K^{(2)\lorentz}_{112;00'}$ and $K^{(2)\lorentz}_{221;00'}$.
Furthermore the remaining lines simplify since $K^{(2)N=4,\lorentz}_{ii;00'}=0$, and one can check that
\be
K^{(2)\lorentz}_{112;00'}=-K^{(2)\lorentz}_{221;00'} = K^{(2)N=4,\lorentz}_{12;00'}-K^{(2)N=4,\lorentz}_{12;0'0}\,.
\ee
In this way all two-loop color structures in the dipole case are expressed in terms of a single angular function.
To evaluate the color factors we recall that while $L_1^aU_1=T^aU_1$, in the antifundamental one has
that $L_2^aU^\dagger_2=-U^\dagger_2T^a$ (this easily follows from $(L_1{+}L_2)U_{12}=0$).
Writing $if^{abc}T^c=[T^a,T^b]$ and collecting terms one easily finds that eq.~(\ref{generalform1}) reduces to
\ba
 \GG^{(2)} \Tr\big[U_1U^\dagger_2\big] &=&
 2\int \frac{d^2\Omega_0}{4\pi} \frac{d^2\Omega_{0'}}{4\pi} K^{(2)N=4,\lorentz}_{12;00'}\left(2U_0^{aa'}U_{0'}^{bb'}-U_{0}^{aa'}U_{0}^{bb'}-U_{0'}^{aa'}U_{0'}^{bb'}\right)
 \nl &&
\hspace{2cm}\times \left(\Tr\big[ [T^{a},T^{b}]U_1 T^{a'}T^{b'}U^\dagger_2\big]+\Tr\big[ T^{b}T^{a} U_1 [T^{b'},T^{a'}]U^\dagger_2\big]\right)
\nl
&&+ \frac{4\pi^2\CA}{3} \int \frac{d^2\Omega_0}{4\pi} \frac{\z{1}{2}}{\z{0}{1}\z{0}{2}}
 \left( \Tr\big[T^{a'} U_1 T^a U^\dagger_2\big]U_0^{a'a}-\CA \Tr\big[U_1U^\dagger_2\big] \right)\,. \label{dipole_nlo}
\ea
This formula, with $K^{(2)N=4,\lorentz}_{12;00'}$ in (\ref{final_result}), is \emph{identical}
to the conformal form of the two-loop evolution obtained by Balitsky and Chirilli, eq.~(6) of ref.~\cite{Balitsky:2009xg}, with
$\frac{\alphas}{4\pi}\GG^{(1)}+\frac{\alphas^2}{16\pi^2}\GG^{(2)}\big|_{\rm here}=-\frac{d}{d\eta}\big|_{\rm there}$, as expected.\footnote{
There is a superficial difference in how we chose to write subtractions, leading to an apparent discrepancy:
\be
2\int \frac{d^2\Omega_0}{4\pi} \frac{d^2\Omega_{0'}}{4\pi}  K^{(2)}_{12;00'} \left( U_{0}^{aa'}U_{0}^{bb'}-U_{0'}^{aa'}U_{0'}^{bb'}\right)\Tr\big[ [T^{a},T^{b}]U_1 [T^{a'},T^{b'}]U^\dagger_2\big]
\propto
\int  \frac{d^2\Omega_{0'}}{4\pi}  \left(K^{(2)}_{12;00'}-K^{(2)}_{21;00'}\right).\nonumber
\ee
That integral however vanishes. This can be easily shown
by noting that being absolutely convergent, the integral defines a Lorentz-covariant function with the same homogeneity in $\b_0,\b_1,\b_2$ as the integrand, hence must a constant times
$\frac{\z{1}{2}}{\z{0}{1}\z{0}{2}}$. The constant vanishes by antisymmetry in  $(\b_1\leftrightarrow \b_2)$.
}
In the planar limit eq.~(\ref{dipole_nlo}) reduces to a closed nonlinear equation for a function of two angles (see
eq.~(\ref{LOplanar})):
\ba
 \GG U_{12} &=&
\left(\frac{\lambda}{16\pi^2}\right)^2 \int\frac{d^2\Omega_0}{4\pi} \frac{d^2\Omega_{0'}}{4\pi}  K^{(2)}_{12;00'}
\left(U_{10}U_{02}+U_{10'}U_{0'2}-2U_{10}U_{00'}U_{0'2}\right)
  \nl && + \frac{\lambda}{8\pi^2}\left(1-\frac{\lambda}{16\pi^2}\frac{\pi^2}{3}\right) \int \frac{d^2\Omega_0}{2\pi}\frac{\z{1}{2}}{\z{0}{1}\z{0}{2}}
 \left(U_{12}-U_{10}U_{02}\right) + O(\lambda^3)\,.
\ea

Going beyond dipoles, rapidity evolution for general products of Wilson lines in the Balitsky-JIMWLK framework has been obtained recently \cite{Balitsky:2013fea,Kovner:2013ona,Kovner:2014xia,Kovner:2014lca},
extending earlier results for two \cite{Balitsky:2008zza,Balitsky:2009xg} and three Wilson lines \cite{Grabovsky:2013mba,Balitsky:2014mca}.
Given the mutual agreement between these works, here we only 
compare directly against the conformal form of ref.~\cite{Kovner:2014xia}. Since the stereographic projection identifies
the $SL(2,\mathbb{C})$ conformal symmetry of the transverse plane with Lorentz symmetry of the two-sphere, this should match with the Lorentz scheme here.

The comparison is in fact straightforward: the range-three kernel $\mathcal{K}_{3,2}$ shown in eq.~(5.12) of ref.~\cite{Kovner:2014xia} is literally
the first four terms of our $K^{(2)\lorentz}_{ijk;00'}$.  The remaining two terms in $K^{(2)\lorentz}_{ijk;00'}$
arise from the telescopic term $F$ in eq.~(\ref{change_in_kernels}) hence do not affect the range-three part.
(These terms are helpful to manifest the convergence at $\b_0\to\b_{0'}$.)
Furthermore, the integral representations for $\mathcal{K}_{3,1}$ and $\mathcal{K}_{3,0}$ in ref.~\cite{Kovner:2014xia}
reproduce the real-virtual pattern embodied in the first line of eq.~(\ref{generalform1}). This demonstrates
the agreement of range-three interactions. Combined with the agreement in the dipole case,
this establishes the complete equivalence of eq.~(\ref{generalform1}) with ref.~\cite{Kovner:2014xia} (and thus, by extension, refs.~\cite{Grabovsky:2013mba,Balitsky:2013fea,Kovner:2013ona}).

In principle, upon linearizing around $U=1$, one also expects complete agreement with the interactions between Reggeized gluons obtained in the BFKL approach.
For two reggeons the agreement was demonstrated at the level of eigenvalues \cite{Fadin:1998py,Ciafaloni:1998gs,Balitsky:2008zza,Balitsky:2009xg}. For three reggeons, it was noted
in ref.~\cite{Grabovsky:2013mba} that a scheme transformation appeared to be missing in order to match with ref.~\cite{Bartels:2012sw}.
This issue should be clarified further.  Here we simply note that there is a natural candidate: the next-to-leading order inner product
(correlator of Wilson lines) \cite{Babansky:2002my,Balitsky:2009yp}.  In the BFKL approach the inner product does not receive loop corrections (the transverse part of the Reggeon propagator remains
$1/p^2$), so only after this effect is removed by a scheme transformation, should agreement be expected.

It is interesting to compare technical aspects of the calculations.
The tree-level soft current (\ref{double_real_tree}) is reminiscent of the light-cone gauge amplitudes in eq.~(43) of ref~\cite{Balitsky:2008zza}.
The subtraction of subdivergences in eq.~(\ref{double_real_master}) is similar to the $+$ prescription derived in refs.~\cite{Balitsky:2001mr,Balitsky:2008zza}.
The transformation to the `Lorentz scheme' (\ref{finite_renorm}) is identical to that leading to the `conformal basis' in refs.~\cite{Balitsky:2009xg,Kovner:2014xia}.
As a significant technical simplification, however, we saved the Fourier transform step.
Also the reliance on standard building blocks made it possible
to benefit from results in the literature, namely the soft currents and collinear splitting functions.

\subsection{Comparison including running coupling}

Having demonstrated the agreement in $\mathcal{N}=4$ SYM, let us now compare the fermion and scalar loop
contributions to the Balitsky-JIMWLK and non-global logarithm Hamiltonians, e.g.
the terms involving $\nF$ and $\nS$ in eq.~(\ref{final_result_matter}).
Performing the comparison with refs.~\cite{Gardi:2006rp,Balitsky:2009xg} we find that the two Hamiltonians agree for the most part, except for the following discrepancy
(setting $z_{ij}=z_i-z_j$):
\ba
 -\frac{d}{d\eta}\big|^{(2)}&=& \GG^{(2)}+
 \betazero \int_{i,j}\!\int \frac{d^2z_0}{\pi} (L_{i;0}^aL_i^a+R_{i;0}^aR_j^a)\left(
   \frac{z_{ij}^2}{z_{0i}^2z_{0j}^2}\log(\mu^2 z_{ij}^2) 
 +\frac{z_{0j}^2-z_{0i}^2}{z_{0i}^2z_{0j}^2}\log\frac{z_{0i}^2}{z_{0j}^2} \right)
 \nl &&
- 2\pi i b_0 \int_{i,j} \log(z_{ij}^2)\big(L_i^aL_j^a-R_i^aR_j^a\big) \label{discrepancy}
\ea
where as before $\mu$ is the $\MSbar$ renormalization scale.
In particular, the difference is proportional to the first $\beta$-function coefficient, as anticipated!
This is very nice since it means it could have been fully reconstructed just by computing a scalar or fermion loop on both sides of the duality.

The origin of the discrepancy (\ref{discrepancy}) is clear: the inversion $y^+\to 1/\mu^2y^+$ in (\ref{conformal}), which relates
the BFKL and non-global log Hamiltonians, is only an isometry up to the Weyl rescaling $ds^2_y \to (\mu y^+)^{-2}ds^2_y$.
This is not a symmetry in a non-conformal theory.  Physically, BFKL and non-global logarithms describe
infinitely fast and infinitely slow measurements of an object's wavefunction, which would not normally be expected
to be connected without conformal symmetry.

For future reference, we note that a general theory deals with Weyl transformations
in non-conformal theories (see for example \cite{Osborn:1991gm}). The essential feature is that, starting from the BFKL side
and performing the conformal transformation (\ref{conformal}), one ends up with
a coordinate-dependent coupling constant:
\be
S'=\int d^4y \frac{-F_{\mu\nu}F^{\mu\nu}}{4\big[g^2(\mu_{0}\mu y^+)\big]},\quad
 \alphas(\mu_{0}\mu y^+) = \alphas(\mu_0)\left(1-2b_0\frac{\alphas(\mu_{0})}{4\pi}\log(\mu y^+)+\ldots\right)\,.
 \label{running_coupling_y}
\ee
In other words, the BFKL Hamiltonian in QCD in principle controls non-global logs in QCD
but in an imagined setup with a coordinate-dependent coupling.
Contrary to real QCD, in this setup a narrow jet never hadronizes:
the increasing coupling due to the growing size of a jet, is compensated by its falloff at large $y^+$.
Thus effectively the coupling is set by the angular size.  This reflects that angles map to distances
in the BFKL problem.
We will not pursue eq.~(\ref{running_coupling_y}) further here,
but in any case it is clear that to all orders in perturbation theory the difference between the BFKL and non-global Hamiltonians will
be proportional to the $\beta$-function (up to scheme transformations).

\section{Higher loops and exponentiation}
\label{sec:higher_orders}

It is instructive to extend the general analysis of section \ref{sec:nlo} to higher loops.
We will (mostly) ignore collinear subdivergences here, concentrating on the soft divergences.

We can organize terms according to the number $m$ of wide-angle partons ($U$ matrices) added to an underlying $n$-jet event.
Our starting point is the known exponentiation of virtual corrections (\ref{factorization_virtual}), which gives the $m=0$ case:
\be
\sigma_0[U] =  \left| \mathcal{P}e^{-\int_0^\mu \Gamma_n} H_{n}(\mu)\right|^2\equiv \mathcal{P}e^{-\int_0^\mu K_0} \sigma^{\ren}_0(\mu)\,.
\ee
The quantity $\sigma^{\ren}_0(\mu)$ is then finite.
For the next case of one wide-angle gluon, a formula was derived in eq.~(\ref{one_loop_master_pre}).
We reproduce it here, in abbreviated notation, omitting $U$ matrices, the angular integration,
$daa^{1-2\eps}$ energy measure, and absolute value squared on the matrix elements:
\be
\sigma_{\leq 1}=
\,\,:\mathcal{P}e^{-\int_0^\mu K_0}
\left[\sigma^{\ren}_0(\mu)+\int\limits_{0<a<\mu} S_1(a;a) H_n(\mu) + \int\limits_{\mu<a<\infty} H_{n{+}1}(a;\mu)\right]\!:
\ee
The colons instruct us to normal-orders terms according to their renormalization scale (largest argument to the right).
As in subsection \ref{ssec:singlereal}, the first integral is identified
as a shift $K_1(\mu)= -S_1(\mu;\mu)$ to the exponent. The remaining (finite) term
then defines the hard coefficient $\sigma^{\ren}_{1}(\mu)$, so that, modulo two real emissions:
\be
\sigma_{\leq 1}= \mathcal{P} e^{-\int_0^\mu(K_0{+}K_1)} (\sigma^{\ren}_{0}(\mu)+\sigma^{\ren}_{1}(\mu))
\equiv \mathcal{P} e^{-\int_0^\mu(K_0{+}K_1)} \sigma^{\ren}_{\leq 1}(\mu)\,.
\ee
Moving on to two real emissions, we follow eq.~(\ref{double_real_regions}) and write the cross-section
as independent emissions plus an additional piece:
\ba
 \sigma_{\leq 2}[U] &=&
 :\mathcal{P}e^{-\int_0^\mu (K_0{+}K_1)}
 \Bigg[ \sigma^{\ren}_{\leq 1}(\mu)
+ \int\limits_{0<a<b<\mu} S^{\rm c}_2(a,b;b) H_n(\mu)
+ \int\limits_{\mu<a<b<\infty} H_{n{+}2}(a,b;\mu)
\nl && \hspace{4cm}
+ \int\limits_{\substack{0<a<\mu \\ \mu<b<\infty}} \big(H_{n{+}2}(a,b;\mu)-S_1(a;a)H_{n{+}1}(b;\mu)\big)
\Bigg]\!: \label{exp2}
\ea
We have introduced the `connected' squared soft current by subtracting all possible subprocesses,
consistent with the energy ordering $a<b<c<\cdots$:
\be
\begin{aligned}
 S_2^{\rm c}(a,b) &= S_2(a,b)-S_1(a)S_1(b)\,,\\
 S_3^{\rm c}(a,b,c) &= S_3(a,b,c)-S_1(a)S_2(b,c)-S_{2}(a,b)S_1(c)+S_1(a)S_2(b)S_3(c)\,, \quad {\rm etc.}
\end{aligned} \label{defcon}
\ee
(In the present abbreviated notation we recall that each factor is a squared soft amplitude, $S_i\equiv |\SS_i|^2$.
Each factor is evaluated at the same renormalization scale, indicated after the semicolon in eq.~(\ref{exp2}).)
Again the first integral in eq.~(\ref{exp2}) is identified as a shift to the exponent,
\be
 K_2(\mu) = -\int_{0<a<\mu} S_2^{\rm c}(a,\mu;\mu),
 \label{Gamma2}
\ee
which generalizes eq.~(\ref{double_real_master}) to include virtual loop effects to all orders.
The (finite) remainder then defines $\sigma^{\ren}_{\leq 2}$.

Using this method it is straightforward to extend the calculation to more radiated particles.
For three radiated particles, for example, after pulling out $\mathcal{P}e^{-\int_0^\mu (K_0{+}K_1{+}K_2)}$ we find again that particles with energy $>\mu$
decouple from divergences, all subdivergences are removed, and
the single divergence gives the shift to the anomalous dimension:
\be
 K_3(\mu) = \,\,\,-\!\!\!\!\!\!\!\int\limits_{0<a<b<\mu} \!\!\!\!\big(S^{\rm c}_3(a,b,\mu;\mu)+K_1(b)\SS^{\rm c}_2(a,\mu;\mu)\big)\,.
\label{Gamma3}
\ee
The second term is present because the exponential of the $K$'s effectively orders the radiation according to the largest momentum
in each connected chunk $K_2$; this over-counts a region where a subsequent emission $K_1$ is harder than the softer parton within $K_2$.
The absence of subdivergences (finiteness of $K_3$ as $\eps\to 0$) in each term is manifest from the fact that
the `connected' squared amplitudes $S^c$ (see eq.~(\ref{defcon}))
vanish near the boundaries $a\to 0$ or $a,b\to 0$. This itself is a consequence
of factorization, or more precisely eq.~(\ref{soft_factorization}) in the form
\be
\lim_{a_1,\ldots, a_k\ll a_{k{+}1},\ldots, a_n} S(a_1,\ldots,a_n;\mu) = S(a_1,\ldots,a_k;\mu) S(a_{k{+}1},\ldots,a_n;\mu)\,. \label{factorization_S}
\ee
It is now clear how to generalize the pattern to higher orders. In fact from the first few cases it appears that a simple formula
gives the anomalous dimension $K$ to all orders:
\be
\framebox{$\displaystyle
K=\Gamma_n + \sum_{k=1}^\infty \,\,\,\int\limits_{a_1<\ldots<a_{k{-}1}<\mu} \!\!\!\overline{\mathcal{P}}e^{\int_{a_1}^\mu K} S^{\rm c}_k(a_1,\ldots,a_{k{-}1},\mu;\mu)\,.
$} \label{all_loop_kernel}
\ee
The exponential factor has a simple physical interpretation as an `exclusion time' effect, and we recall that $a$'s are the energies of real radiated particles.  We have verified explicitly (with the help of a computer) that exponentiating $K$ using eq.~(\ref{factorization}) reproduces all contributions where up to at least 9 real particles have energy below $\mu$, so we believe that the formula is correct to all orders.

Equation (\ref{all_loop_kernel}) is one of the main results of this paper.
It expresses, to all loop orders, the Hamiltonian governing non-global logarithms
as a convergent integral over finite, well-defined building blocks,
generalizing the eqs.~(\ref{double_real_master}) and (\ref{one_loop_master}) used in the two-loop computation.
The building blocks are the squares of the
infrared-renormalized soft currents (which include virtual loops to all orders), defined in eq.~(\ref{soft_factorization}).
Only the $\eps^0$ part of the infrared-renormalized currents are needed, in agreement with the general arguments of ref.~\cite{Weinzierl:2011uz}.

Since the exponent $K$ is manifestly finite as $\eps\to 0$
(being expressed in terms of connected squared soft currents),
the formula also demonstrates to all loops that infrared divergences exponentiate according to eq.~(\ref{factorization}).
The physical inputs were the known exponentiation (\ref{factorization_virtual}) of virtual corrections, plus the factorization of successive real emissions (\ref{factorization_S}); eq.~(\ref{factorization}) comes out as a purely combinatorial output.

To fully prove eq.~(\ref{factorization}) one should address the issue of collinear subdivergences, omitted in the present discussion.
Physically we expect these to cancel, since the operator definition of $\sigma[U]$ is collinear-safe.
In subsection \ref{ssec:singlereal} this was made manifest by defining collinear-subtracted real and virtual
contributions, such that their sum was unaffected by the subtraction.
We have no reason to think that this couldn't be achieved at higher orders as well,
following the method of ref.~\cite{Catani:1999ss}.

We should mention that eq.~(\ref{Gamma3}) gives the evolution equation in a non-minimal scheme:
the two-loop exponent $K^{(2)}_2$ in eq.~(\ref{Gamma2}) depends on $\eps$ through a factor $a^{-2\eps}$
and thus differs from the $\MSbar$ result of this paper by terms proportional to $\eps$.
These are interpretable as a renormalization which shifts $\GG^{(3)}$ by a finite commutator, giving, instead of eq.~(\ref{Gamma3}):
\be
 K_3^{(3),\MSbar} =
 \,\,\,-\!\!\!\!\!\!\!\int\limits_{0<a<b<\mu} \!\!\!\!\!\!S^{\rm c}_3(a,b,\mu) \,\,\,-\!\!\!\!\int\limits_{0<a<\mu} 
 \frac12\left(K^{(1)}_1S^{\rm c}_2(a,\mu)+S^{\rm c}_2(a,\mu)K^{(1)}_1\right)\log\frac{\mu}{a} + O(\eps)\,. \label{Gamma3a}
\ee
Discrepancies in $K_1^{(2)}$ at $O(\eps)$ add other commutator terms.
All these are unrelated to a further finite renormalization needed to make Lorentz covariance manifest.
As in eq.~(\ref{scheme_eps}) it can be fully predicted
by upgrading the two-loop result (\ref{final_result}) to a $D$-dimensional covariant form.
Although these finite renormalizations become combinatorially very complicated at higher loop orders, being finite they
cannot interfere with the statement (\ref{factorization}) of exponentiation.

Finally, we list the ingredients needed to evaluate eq.~(\ref{all_loop_kernel}) at three-loop:
\begin{itemize}
\item The tree-level soft current for three soft gluons $\SS_3^{(0)}$,
at one-loop for two gluons $\SS_2^{(1)}$,
and two-loop for one gluon $\SS_1^{(2)}$,
and the three-loop soft anomalous dimension $\Gamma^{(3)}_n$.
\item The next-to-leading order $1\to 2$ and tree-level $1\to 3$ collinear splitting amplitudes \cite{Bern:1999ry}.
\end{itemize}
The two-loop soft current and three-loop soft anomalous dimension
are presently known for two hard partons \cite{Moch:2005tm,Duhr:2013msa,Li:2013lsa}.
Unfortunately this will not suffice for non-global logarithms nor BFKL in general, since each radiated gluon
counts like a hard one from the point of view of softer radiation.\footnote{
An interesting possibility is that other constraints, such as the KMS condition; Lorentz symmetry ($SL(2,\mathbb{C})$); collinear singularities;
CPT symmetry; and Hermiticity of the BFKL Hamiltonian could uniquely fix $K^{(3)}$ without these building blocks, effectively determining them.}
However, for dipole evolution in the planar limit, everything needed is already known.

\section{Conclusion}
\label{sec:conclusion}

In this paper we considered a `color density matrix' which aims to characterize soft radiation
in gauge theory. Particles have the same four-momenta on each side of the density matrix, but different colors.
We argued that it should resum large logarithms arising in the presence of wide-angle
phase space cutoffs, so-called non-global logarithms, to all orders in logarithms and $1/N_c$.
We proved the all-order exponentiation of infrared divergences for this object in terms of an anomalous dimension
$K$ (see eq.~(\ref{factorization})), constructed formally in eq.~(\ref{all_loop_kernel}), modulo
the technical assumption that collinear subdivergences cancel.
We explicitly computed $K$ to two-loop (eqs.~(\ref{generalform1})--(\ref{final_result_matter})) and performed a number of checks on this result.
We also stressed the equality between $K$ and the BFKL Hamiltonian, which allows our results to be viewed as an independent derivation of the next-to-leading order BFKL Hamiltonian, obtained here directly in a novel, compact form.

The procedure to calculate a cross-section receiving non-global logarithms was sketched in the introduction.
One distinguishes infrared and ultraviolet scales, which are to be connected by evolving using $K$.
At both ends lie finite quantities: an `IR measurement' which contains
details of the experimental definition of a `soft' particle, and corresponding vetoes;
an `UV measurement' which depends on the initial state and possible vetoes imposing hard jets in the final state.
The logic of factorization being that their calculations are independent of each other,
we focused in this paper on the (universal) evolution $K$.
Study of the infrared-finite, but process-dependent, measurement functions is left to future work,
for example matching with the fixed-order results \cite{Kelley:2011ng}, as well as phenomenological studies.

Mathematically, $K$ is an integro-differential operator acting on functionals $\sigma[U]$
of a two-dimensional field of unitary matrices $U(\theta)$ (e.g. $SU(3)$ matrices in QCD),
with $\theta$ an angle in the detector.
This means that $K$ cannot be diagonalized explicitly.
Although it is a quite complicated object, it is a useful starting point for further approximations.
These include, as reviewed in section \ref{sec:review},
numerical Monte-Carlo techniques at finite $N_c$, reduction to an ordinary integro-differential equation at large $N_c$,
or linearization {\it \`a la} BFKL around $U=1$.  We hope that the compact form of next-to-leading order evolution
obtained in this paper (eqs.~(\ref{generalform1}) and below) will prove convenient for a next-to-leading order numerical implementation.

For application to hadron colliders it will be important to go beyond the limitation of an initial color-singlet object,
as done in this paper, and allow for initial state radiation.
This could lead to additional (super-leading? \cite{Forshaw:2006fk,Keates:2009dn}) effects related to subtle color-dependent phases in collinear limits \cite{Catani:2011st,Forshaw:2012bi}.

The formalism does not distinguish between global and non-global logarithms, but it is easy to see how it simplifies
in the case of global observables.
For example, when radiation is excluded everywhere but inside narrow cones, the IR averaging procedure
sets $\l U\r=0$ outside these cones which effectively shuts down the real terms in the evolution. It is then dominated by virtual effects,
as is usual for global observables.
It is only for observables sensitive to details of wide-angle radiation that the complications of the formalism kick in.
It would be interesting to connect the present approach with that of ref.~\cite{Banfi:2014sua},
which deals with recursive infrared and collinear safe event shapes (`rIRC').

There has been recent activity regarding formal aspects of measurements at infinity,
in connection for example with the Bondi, van der Burg, Metzner and Sachs (BMS) symmetry \cite{Georgi:2014sxa,Strominger:2013lka}.
The density matrix construction could be useful in this context.

From a theoretical perspective, the Hamiltonian $K$ connects, in a unified way, the following gauge-theory concepts: 
the cusp anomalous dimension (governing global logarithms);
the KLN theorem (cancelation of collinear and infrared divergences);
the factorization of soft radiation;
the BFKL equation.

The equivalence with BFKL, verified explicitly in section \ref{sec:bfkl}, 
is a consequence of conformal symmetry \cite{Hatta:2008st} and is an equality up to $\beta$-function terms
(fixed by comparatively simpler matter loops (\ref{discrepancy})).
The basic physical intuition is summarized in fig.~\ref{fig:bfkl}.
Remarkably, properties manifest in one context are not necessarily so in the other.

For example, one fundamental assumption in both the BFKL and Balitsky-JIMWLK frameworks
is that transverse integrals should be saturated by transverse scales that do not grow linearly with $s$,
$\sim t\ll s$, ensuring that rapidity logarithms ($\log s$) arise only from longitudinal integrations \cite{Balitsky:1995ub,Lipatov:1995pn}.
While reasonable it is unclear how one would prove this directly
beyond the current state of the art, e.g. next-to-leading log. The correspondence with non-global logarithms immediately implies it to all orders, since it amounts to the amply understood cancelation of collinear divergences.  The non-global logarithm formulation
also seems to be computationally advantageous, as discussed in sections \ref{ssec:N4comp} and \ref{sec:higher_orders}.

In the other direction, the phenomenon of gluon Reggeization
suggested a compact way to write the evolution equation (see eq.~(\ref{generalform})),
which manifests a connection between real and virtual effects.
Intriguingly, we found that these relations could perhaps also be explained by the Feynman tree theorem.
It would be very interesting to see if either of these approaches generalizes to higher loop orders.

Finally, we mention that the simplest non-global logarithms to resum in this framework (beyond the planar limit)
involve situations close to the linear regime $U\approx 1$, where
the linearized equation has lowest eigenvalue the well-known Pomeron intercept $-\frac{4\alphas \CA\log 2}{\pi}$.
Naively this regime might correspond to multiplicity-type measurements, e.g.
counting away jet charged tracks as a function of angle and an energy cutoff.
Perhaps this or some other observables could provide an indirect experimental handle on the BFKL Pomeron.

\acknowledgments{I would like to thank Jacob Bourjaily, Ilya Feige and Matthew Schwartz for stimulating discussions,
and Michael Trott and Alberto Guffanti for a well-timed conversation on the topic of $\pi^2$ resummation.}

\begin{appendix}

\section{Single-real and double-virtual contributions}
\label{app:virtual}

We detail our evaluation of eq.~(\ref{one_loop_master}).
The starting point is the one-loop soft current (for emitting one soft gluon), reproduced in eq.~(\ref{SS1_virtual_colors}).
We need to convert it to the barred scheme,
\be
\bar{\SS}^{(1)}_1= \SS^{(1)}_1 +
\left[\bar{X}^{(1)},\SS^{(0)}_1\right]
 \qquad\mbox{where}\qquad \bar{H}^{(1)}_n-H^{(1)}_n\equiv \bar{X}^{(1)}H^{(0)}_n\,,
\label{definitionofX}
\ee
where $\bar{H}$ implements the subtraction in eq.~(\ref{primedscheme}) of collinear splitting functions.
Since the splitting functions for all but the radiated gluon cancel in the commutator,
we will only need the gluon splitting function
\be\begin{aligned}
\Splitsq{g}{a\beta_0,b\beta_{0'}} &=\frac{\CA(b-a)^{-2\eps}}{ab(b^{-2\eps})}\left(
\frac{2}{x(1{-}x)\alpha_{00'}}{+}\frac{(n_{\rm Weyl}^{\adj}-4)}{\alpha_{00'}}
{+}x(1{-}x)(n_s^{\rm adj}{-}2n_{\rm Weyl}^{\adj}{+}2) f\right),
\end{aligned}\label{split}\ee
where $x=a/b$.
The prefactor has a kinematical origin and accounts for the change in the measure $b^{1{-}2\eps}db$.
The computation of such functions is standard \cite{Catani:1996vz}.
In the $x$-dependence one can recognize various DGLAP kernels $P_{g\to(\cdots)}(x)$, as expected.
We use the dimensional reduction scheme so the parenthesis does not depend on $\eps$.
(Regarding color factors we recall that we show intermediate formulas only in a theory with color-adjoint matter.)
The scalar contribution to the splitting function is polarization-dependent and for us the most useful information
will be its dot product against $\b_i^\mu \b_j^\nu$, divided by $\b_i{\cdot}\b_j$: this is what enters
eq.~(\ref{one_loop_master}). This is given by
\be\begin{aligned}
f(\b_0,\b_{0'};\b_i,\b_j) &= \frac{\big[\z{0}{i}-\z{0'}{i}\big]\big[\z{0}{j}-\z{0'}{j}\big]}{\z{i}{j}\z{0}{0'}^2}
\\ &= \frac{\big[\z{0'}{i}(\z{0}{j}-\z{0}{0'})-\z{0'}{j}(\z{0}{i}-\z{0}{0'})\big]^2}{2\z{i}{j}\zsq{0}{0'}\z{0'}{i}\z{0'}{j}} + \mbox{convergent or telescopic}\,.
\end{aligned}\ee
The first form is obtained directly from the Feynman rules and
makes manifest that the dependence on $\b_i,\b_j$ is consistent with factorization.
We will prefer the second form, which provides a closer match with eq.~(\ref{realcorrections}) and also yields a simpler integrated expression.
It differs by terms which are either convergent or vanish using color conservation.
Computing the integral in (\ref{primedscheme}) we then obtain
\be
\bar{X}^{(1)}=-\sum_{i,j}R_i^aR_j^a \big(L_2(\z{i}{j}){+}\log4\log \z{i}{j}\big)
+\sum_{i=g}\left(-\frac{\pi^2\CA}{3}+\gammaK^{(2)}
+\betazero\log\frac{\mu}{2k_i^0}{+}\frac{\CA}2\log^2\frac{\mu}{2k_i^0}\right),
\label{Xexpression}\ee
with $L_2(x)\equiv \Li_2(1-x)+\frac12\log^2(x) -\frac{\pi^2}{6}$.
The sum runs over gluons to stress that we haven't computed the other cases,
and the cusp anomalous dimension is in eq.~(\ref{H2_bare_version}).

The commutator then easily yields
\begin{subequations}
\ba
\bar{S}^{(1)\mu}_i(k) &=& \left[
\gammaK^{(2)}
+(\betazero-2\pi i\CA)\log\frac{\mu}{2k^0} \right]\frac{\beta_i^\mu}{\beta_i{\cdot}k}
\label{Soneprimea}\\
\bar{S}^{(1)\mu}_{ij}(k) &=&  \left( \frac{\b_j^\mu}{\b_j{\cdot}\b_0}-\frac{\b_i^\mu}{\b_i{\cdot}\b_0}\right)
 \left(\frac12\log^2\left(\frac{\z{0}{i}\z{0}{j}}{\z{i}{j}}\right) {+}L_2(\z{i}{j}){-}L_2(\z{0}{i})-L_2(\z{0}{j})
\right.\nl &&\hspace{4.0cm}\left.
+\log \frac{e^{-i\pi}k_0^2}{\mu^2}\log\frac{\z{0}{i}\z{0}{j}}{\z{i}{j}}
  \right).
\ea\label{Soneprime}\end{subequations}
We stress that only the $O(\eps^0)$ terms of $\SS_1^{(1)}$ were needed to obtain this. It is noteworthy
that
the $-\pi^2\CA/6$ from the original soft function, the $-\pi^2\CA/3$ from the scheme transformation,
and the $-\CA\log(e^{-i\pi})^2/2$ from the phase of the logarithm have nicely canceled to leave the cusp anomalous dimension.

Substituting into eq.~(\ref{one_loop_master}) the soft factor produces two color structures
\ba
 \GG^{(2)}\big|_{{\rm linear\,in\,U}} &=&
\phantom{+} \sum_{i,j,k}\int\frac{d^2\Omega_0}{4\pi}G^{(2)}_{ijk;0} if^{abc}\big(L_i^{a'}U_0^{a'a}R_j^bR_k^c - L_j^b L_k^c U_0^{aa'}R_i^c\big)
 \nl && +\sum_{i,j}\int\frac{d^2\Omega_0}{4\pi}G^{(2)}_{ij;0} U_0^{aa'}(L_i^aR_j^{a'}+R_i^{a'}L_j^{a})\label{colors_virtual}\,,
\ea
which multiply the angular functions
\be
G^{(2)}_{ijk;0}=-4S^{\mu}_i(\b_0)\bar{S}^{(1)\mu}_{jk}(\b_0)\big|_{k^0=\mu}\,,
\qquad
G^{(2)}_{ij;0}=-4S^{(0)\mu}_i \bar{S}^{(1)\mu}_{j}(\b_0)\big|_{k^0=\mu}\,. \label{result_single_real}
\ee
These can be evaluated explicitly using (\ref{Soneprime}). The remaining linear-in-$U$ contribution, the subtraction (\ref{double_real_master2}), is simply
\be\begin{aligned}
\GG^{(2)}_{\rm sub.} &=
{-}\sum_{i,j} \int \frac{d^2\Omega_0}{4\pi}\frac{d^2\Omega_{0'}}{4\pi} \frac{\z{i}{j}\CA}{\z{0'}{i}\z{0'}{j}}U_{0'}^{aa'} (L_i^aR_j^{a'}{+}R_i^{a'}L_j^a)
\left[\frac{n_{\rm Weyl}^{\adj}{-}4}{\z{0}{0'}} {+}\frac{1}{6}\big(n_s^{\adj}{-}2n_{\rm Weyl}^{\adj}{+}2\big)f\right]\!.
\end{aligned}\label{subtract_linear}\ee
This is to be added at the integrand level to eq.~(\ref{realcorrections}) and removes its collinear divergences.
(Since the cancellation occurs at the integrand level in eq.~(\ref{double_real_master}), it is justified to set $\eps=0$ in both.)

The result (\ref{result_single_real})-(\ref{subtract_linear}) is now to be compared against the prediction from the ansatz (\ref{generalform}),
which gives the same color structures and predicts the first angular functions as
\be\begin{aligned}
 G^{(2)}_{ijk;0}\big|_{\rm ansatz} &= \int \frac{d^2\Omega_{0'}}{4\pi} \left(K^{(2)}_{ikj;00'}-K^{(2)}_{ijk;00'}\right)
 \\ &=
 2\left(\frac{\z{i}{k}}{\z{0}{i}\z{0}{k}}-\frac{\z{i}{j}}{\z{0}{i}\z{0}{j}}\right)
\left(\frac12\log^2\left(\frac{\z{0}{j}\z{0}{k}}{\z{j}{k}}\right)+L_2(\z{j}{k})-L_2(\z{0}{j})-L_2(\z{0}{k})\right)\,. \label{ansatz_angular_function1}
\end{aligned}\ee
This agrees precisely with eq.~(\ref{Soneprimea}), up to the $i\pi$ term recorded in eq.~(\ref{imaginary_term}).
For the other structure
\ba
 G^{(2)}_{ij;0'}\big|_{\rm actual}-\big|_{\rm ansatz} &=&
\frac{\z{i}{j}}{\z{0}{i}\z{0}{j}}\left[2\gammaK^{(2)}-b_0\log 4\right]- K^{(2)}_{ij;0'}+ {\rm eq.~(\ref{subtract_linear})}
+ \CA\int \frac{d^2\Omega_{0}}{4\pi} K^{(2)}_{ij;00'}
  \nl
 &=&\frac{\z{i}{j}}{\z{0}{i}\z{0}{j}}\left[\gammaK^{(2)} +\betazero\log\frac{\z{i}{j}}{4\z{0}{i}\z{0}{j}} \right] -K^{(2)}_{ij;0},
\ea
which fixes $K^{(2)}_{ij;0'}$ as recorded in the main text.

Finally we check the double-virtual terms.  To get the prediction from the ansatz (\ref{generalform})
we need to integrate (\ref{ansatz_angular_function1}).
The $L_2(\z{0}{k})$ terms look scary, but they cancel out trivially because one needs only
the total antisymmetrization of $G_{ijk;00'}^{(2)}$ modulo terms which do not depend on all three labels simultaneously.
The integral is still a bit nontrivial but we could simplify its antisymmetric part using integration-by-parts.
We omit the details and quote only the rather simple result for the $G_{ijk;0}^{(2)}$ contribution,
\be
if^{abc}R_i^aR_j^bR_k^c\sum_{i,j,k}\int \frac{d^2\Omega_0}{4\pi}\left(-\frac12G_{ijk;0}^{(2)}\right) =
8if^{abc}\sum_{i,j,k}R_i^aR_j^bR_k^c \log(\z{i}{j})L_2(\z{j}{k})\,.
\ee
Finally the other term in the ansatz is (dropping terms depending on one particle at a time)
\be
-\sum_{i,j}R_i^aR_j^a \int\frac{d^2\Omega_0}{4\pi}\frac{\z{i}{j}}{\z{0}{i}\z{0}{j}}
K^{(2)}_{ij;0}
\simeq -2\sum_{i\neq j}R_i^aR_j^a \left(\gamma_K^{(2)}\log(\z{i}{j})-\betazero(L_2\big(\z{i}{j})+\log4\log\z{i}{j}\big)\right).\nonumber
\ee
The preceding two equations are easily verified to be in perfect agreement with the commutator (\ref{Gamma2_prime}), proving that the ansatz does not miss any double-virtual term.

As a final comment, we note that the $L_2$ function and most $\log 2$'s have a simple origin: the scheme change (\ref{finite_renorm}).
For example $\int\frac{d^{2{-}2\eps}\Omega_0}{4\pi(2\pi)^{-\eps}c_\Gamma}\delta^{(1)}_{ij;0}=-\frac{1}{\eps^2}+L_2(\z{i}{j})+\log4\log(2\z{i}{j})+O(\eps)$.
With hindsight, we could have saved ourselves much algebra by switching from the $\MSbar$ to the Lorentz-covariant
scheme from the very beginning, which would have prevented $L_2$ and most $\log 2$'s from ever appearing.


\end{appendix}

\bibliographystyle{JHEP}
\bibliography{ngl}

\end{document}